\documentclass[12pt,pdftex,preprint]{aastex}
\usepackage{amssymb,amsmath,mathrsfs}

\newlength{\figurewidth}
\newcommand{\foreign}[1]{\textit{#1}}
\newcommand{\etal}{\foreign{et\,al.}}
\newcommand{\documentname}{\textsl{Note}}
\newcommand{\project}[1]{\textsl{#1}}
\newcommand{\thetractor}{\project{The~Tractor}}
\newcommand{\sdss}{\project{SDSS}}

\newcommand{\tmatrix}[1]{\boldsymbol{#1}}
\newcommand{\inverse}[1]{{#1}^{-1}}
\newcommand{\transpose}[1]{{#1}^{\mathsf T}}
\newcommand{\tvector}[1]{\boldsymbol{#1}}
\newcommand{\pos}{\tvector{x}}
\newcommand{\spos}{\tvector{\xi}}
\newcommand{\mean}{\tvector{m}}
\newcommand{\var}{\tmatrix{V}}
\newcommand{\affine}{\tmatrix{R}}
\newcommand{\uv}{\tvector{u}}
\newcommand{\zero}{\tmatrix{0}}
\newcommand{\identity}{\tmatrix{I}}
\newcommand{\normal}{N}
\newcommand{\given}{\,|\,}
\renewcommand{\star}{\mathrm{star}}
\newcommand{\dev}{\mathrm{dev}}
\newcommand{\ser}{\mathrm{ser}}
\newcommand{\lux}{\mathrm{lux}}
\newcommand{\luv}{\mathrm{luv}}

\newlength{\figwidth}
\begin{document}

\title{Replacing standard galaxy profiles with \\ mixtures of Gaussians}
\author{David W. Hogg\altaffilmark{1,2,3} \&
        Dustin Lang\altaffilmark{4,5}}
\altaffiltext{1}{To whom correspondence should be addressed; \texttt{david.hogg@nyu.edu}}
\altaffiltext{2}{Center for Cosmology and Particle Physics, Department of Physics, New York University, 4 Washington Place, New York, NY 10003, USA}
\altaffiltext{3}{Max-Planck-Institut f\"ur Astronomie, K\"onigstuhl 17, D-69117 Heidelberg, Germany}
\altaffiltext{4}{Princeton University Observatory, Princeton, NJ 08544, USA}
\altaffiltext{5}{McWilliams Center for Cosmology, Carnegie Mellon University, 5000 Forbes Avenue, Pittsburgh, PA 15213, USA}

\begin{abstract}
Exponential, de~Vaucouleurs, and S\'ersic profiles are simple and
successful models for fitting two-dimensional images of galaxies.  One
numerical issue encountered in this kind of fitting is the pixel
rendering and convolution (or correlation) of the models with the
telescope point-spread function (PSF); these operations are slow, and
easy to get slightly wrong at small radii.  Here we exploit the
realization that these models can be approximated to arbitrary
accuracy with a mixture (linear superposition) of two-dimensional
Gaussians (MoGs).  MoGs are fast to render and fast to
affine-transform.  Most importantly, if you have a MoG model for the
pixel-convolved PSF, the PSF-convolved, affine-transformed galaxy
models are themselves MoGs and therefore very fast to compute,
integrate, and render precisely.  We present worked examples that can
be directly used in image fitting; we are using them ourselves.  The
MoG profiles we provide can be swapped in to replace the standard
models in any image-fitting code; they sped up model fitting in our
projects by an order of magnitude; they ought to make any code faster
at essentially no cost in precision.
\end{abstract}

Gaussians are remarkable distribution functions.  They have the
incredible properties that---in any number of dimensions---the
convolution (or correlation) of one multivariate Gaussian with another
is itself a multivariate Gaussian, and any product of multivariate
Gaussians is itself a multivariate Gaussian but with a different
normalization.  Furthermore, the means and variance tensors of the
Gaussians output by these operations are related simply to the means
and variance tensors of the inputs.  Add to these wonders the fact
that Gaussians form a complete basis for representing (smooth)
probability distribution functions and it becomes remarkable that we
don't do everything we do in terms of Gaussians.

To elaborate, a mixture of multivariate Gaussians---a linear
superposition---can be used to represent any reasonable distribution
in any number of dimensions to any reasonable precision.  Convolution
(or correlation) by any other distribution that has also been
represented by a mixture of multivariate Gaussians creates a new
mixture of Gaussians with simply adjusted amplitudes, means, and
variance tensors.  The ubiquity of convolution operations in astronomy
suggests the widespread adoption of mixture-of-Gaussian (MoG)
modeling.  We have pioneered this in the area of distribution modeling
in large numbers of dimensions (\citealt{xd, xdqso, xdqsoz}), where
convolution occurs because the true or noise-free distribution is
convolved with the noise before being observed.  Here we are going to
capitalize on the convolution properties of MoGs in modeling galaxy
morphologies in imaging data, where the true or
high-angular-resolution intensity field is convolved with the
point-spread function (PSF) before being observed.  This is always
easier with Gaussians than cuspy profiles, but is \emph{particularly}
useful in data sets in which the PSF itself has also been modeled as a
MoG, which is not uncommon (for example, the \project{Sloan Digital
  Sky Survey} imaging pipelines described by \citealt{lupton} make
approximate MoG PSF models for every imaging field).

We are not the first in this space: Deconvolution and modeling of
galaxy images with MoGs has been done very successfully before (for
example, \citealt{bendinelli, emsellem, bendinelli2, cappellari}; they
use the name ``multi-Gaussian expansion'' or MGE).  However, the idea
behind those projects was to use the MoGs to provide a very free form
for the isophotes or morphologies of resolved galaxies or other
complex scenes.  Here our goals are very limited: We want to improve
the performance of standard galaxy image model fitting by expressing
the standard galaxy models---the exponential and de~Vaucouleurs
profiles---as rigid MoGs.

Whenever an investigator is fitting PSF-convolved exponential or
de~Vaucouleurs profiles, the models presented here will improve code
performance.  That doesn't mean that doing such fitting is a Good
Idea.  These profiles are effective models for galaxies at low
signal-to-noise (as they are used in the \project{SDSS}) and they are
useful templates for performing consistent photometry across
morphologically diverse objects (as we use them in \citealt{bundy}).
More radically, these profiles are sometimes used to distinguish disk
and bulge light (as in \citealt{bulgedisk} and references cited
therein); these uses are prone to over-interpretation: There is no
theoretical argument and only weak observational arguments that the
rotation-supported parts of galaxies are always exponential and that
the kinematically hot components are always more de-Vaucouleurs-like.

The performance advantages we will obtain from using MoG
approximations are not simply that convolution itself is trivial.  The
standard unconvolved galaxy models, especially the de~Vaucouleurs and
S\'ersic models (\citealt{dev}, \citealt{ser}), are very ill-behaved
near the galaxy center.  Rendering these profiles precisely near the
center can be very challenging numerically.  In addition, most of the
rendering time is spent at very small radii, where PSF convolution is
going to erase all structure anyway.  That is, a MoG description of
the de~Vaucouleurs profile saves time both in rendering \emph{and} in
PSF convolution; it produces profiles that are approximate but very
high performance when all real uses of the profiles are PSF-convolved,
as they usually are.

On the subject of PSF-convolution, it is important in any
image-modeling situation to think of the PSF as the
\emph{pixel-convolved} point-spread function.  Under this choice,
synthesis of a pixelized image involves only convolution with the PSF
and evaluation at the pixel centers.  That is, image synthesis
(returned to at the end of this \documentname) consists of PSF
convolution---an arithmetic operation on the two MoGs---followed by
evaluation of the resulting MoG at the pixel centers).

We need good performance in two-dimensional image synthesis (for
fitting) because, with \thetractor\ (Lang \etal, forthcoming), we are
building a comprehensive model of all the imaging data we have; this
will have models of many millions of galaxies in imaging that contains
on the order of $10^{13}$ pixels.  We are basing our models on the
\project{Sloan Digital Sky Survey} (\sdss) Catalog galaxy models,
which only include exponential, de~Vaucouleurs, and composite (mixture
of the two) radial profiles.  In detail, in fact, the \sdss\ Catalog
models are small modifications of these; details below.  So our goal
here is to provide replacements for these models in order to improve
the performance of galaxy image modeling and analysis software.  Some
of the models we present have been used previously in tools for
precise photometry (\citealt{bundy}), and are being used in the Large
Synoptic Survey Telescope (LSST) prototype galaxy photometry pipeline
(\citealt{shaw}), in which these models are convolved with a shapelet
representation of the PSF; because shapelets are simply polynomial
perturbations on a Gaussian, they can also be convolved analytically
with mixture-of-Gaussian galaxy models (\citealt{bosch}).  These
models could also be used to speed other image-fitting systems, like
the very successful \project{GALFIT} (\citealt{galfit}).  An
alternative to making MoG profiles is to find exact analytic
expressions for certain kinds of convolutions.  There \emph{are} some
analytic results for PSF-convolved S\'ersic models but they involve
some non-trivial series and special-function expressions
(\citealt{trujillo}); we won't come back to these again but they might
be very useful in many situations.

Because we are thinking about two-dimensional imaging, we use here
two-dimensional Gaussian or Normal distributions, which look like
\begin{eqnarray}\displaystyle
\normal(\pos\given\mean,\var) &\equiv& \frac{1}{2\pi}\,\det(\var)^{-1/2}\,\exp(-\frac{1}{2}\,\transpose{[\pos-\mean]}\cdot\inverse{\var}\cdot[\pos-\mean])
\quad ,
\end{eqnarray}
where $\pos$ and $\mean$ are two-dimensional vectors (usually in the
focal plane or on the sky or something like that), and $\var$ is a
symmetric $2\times 2$ variance tensor or matrix, and implicitly the
vectors are column vectors.  A mixture of Gaussians is a linear
superposition of Gaussians.  Any positive two-dimensional function
with finite support and finite total integral---including as a special
case any two-dimensional probability distribution function---can be
represented as a MoG to arbitrary accuracy; that is
\begin{eqnarray}
p(\pos) &\approx& \sum_{k=1}^K a_k\,\normal(\pos\given\mean_k,\var_k)
\\
1 &=& \sum_{k=1}^K a_k
\quad ,
\end{eqnarray}
where $p(\pos)$ is any probability distribution of a two-dimensional
quantity $\pos$, the $\approx$ symbol implies approximation, $K$ is
the number of Gaussians used in the MoG, and the $K$ Gaussians
have amplitudes $a_k$, means $\mean_k$, and variance tensors $\var_k$.
The sum-to-one condition ensures that the probability distribution
approximation is properly normalized.

We wish to make an approximation to the two-dimensional circular
exponential (exp) profile $Q^{\exp}(\cdot)$
\begin{eqnarray}\displaystyle
Q^{\exp}(\spos) &\equiv& \exp(-\alpha^{\exp}\,[|\spos| - 1])
\\
\alpha^{\exp} &\equiv& 1.67834699
\quad ,
\end{eqnarray}
where $\spos$ is a dimensionless focal-plane position, and
$\alpha^{\exp}$ is a dimensionless inverse length set to ensure that
the profile has unit half-light radius.  The position $\spos$ is
dimensionless because it parameterizes the unit-size dimensionless
function.  We seek the best (where ``best'' will be defined below)
$M^{\exp}$-Gaussian MoG (where $M^{\exp}$ is an integer) approximation
\begin{eqnarray}\displaystyle
Q^{\exp}(\spos) &\approx& \sum_{m=1}^{M^{\exp}} a^{\exp}_m\,\normal(\spos\given\zero,\var^{\exp}_m)
\\
\var^{\exp}_m &\equiv& v^{\exp}_m\,\identity
\quad ,
\end{eqnarray}
where all of the means are exactly zero and all of the variances
$\var^{\exp}_m$ in the MoG can be represented as a scalar
$v^{\exp}_m$ multiplied by the identity matrix $\identity$
because we are requiring this dimensionless function to be precisely
circular (so every component is itself circular and concentric).
Similarly for the de~Vaucouleurs (dev) profile
\begin{eqnarray}\displaystyle
Q^{\dev}(\spos) &\equiv& \exp(-\alpha^{\dev}\,[|\spos|^{1/4} - 1])
\\
\alpha^{\dev} &\equiv& 7.66924944
\\
Q^{\dev}(\spos) &\approx& \sum_{m=1}^{M^{\dev}} a^{\dev}_m\,\normal(\spos\given\zero,\var^{\dev}_m)
\\
\var^{\dev}_m &\equiv& v^{\dev}_m\,\identity
\quad .
\end{eqnarray}
The half-light inverse-radius parameters $\alpha^{\exp}$ and
$\alpha^{\dev}$ are from \citet{ciotti}.  The challenge we
meet below is to determine the parameters
\begin{eqnarray}
\{a^{\exp}_m,v^{\exp}_m\}_{m=1}^{M^{\exp}}~,~\{a^{\dev}_m,v^{\dev}_m\}_{m=1}^{M^{\dev}}
\end{eqnarray}
to best approximate the traditional galaxy profile functions, under
some sensible definition of the word ``best'', as a function of the
model complexity parameters (numbers of components) $(M^{\exp},
M^{\dev})$.

In addition to these, there are general S\'ersic (``ser'') profiles, of
which the exp and dev profiles are special cases.  The general ser
profile has one parameter (the ``index'') $n$:
\begin{eqnarray}\displaystyle
Q^{\ser(n)}(\spos) &\equiv& \exp(-\alpha^{\ser(n)}\,[|\spos|^{1/n} - 1])
\\
\{\alpha^{\ser(2)}, \alpha^{\ser(3)}, \alpha^{\ser(5)}\} &\equiv& \{3.67206075, 5.67016119, 9.66871461\}
\quad ,
\end{eqnarray}
where we have given the constant $\alpha^{\ser(n)}$ for just a few
values of $n$ (\citealt{ciotti}; the $\exp$ and $\dev$ profiles given
above provide values for $n=1$ and $n=4$.)

The \sdss\ pipelines (\citealt{lupton}) make use of modified profiles,
which have been truncated smoothly at large radius and (in the case of
the de~Vaucouleurs profile) ``softened'' at the center.  The
\sdss\ form of the exponential (lux) profile is
\begin{eqnarray}\displaystyle
Q^{\lux}(\spos) &\equiv& \left\{\begin{array}{ll}
  \exp(-\alpha^{\lux}\,[|\spos| - 1]) & \mbox{for~}|\spos| < 3 \\
  \exp(-\alpha^{\lux}\,[|\spos| - 1])
  \,\left[1 - [|\spos| - 3]^2\right]^2 & \mbox{for~}3 < |\spos| < 4 \\
  0                                   & \mbox{for~}4 < |\spos|
\end{array}\right.
\\
\alpha^{\lux} &\equiv& 1.67835
\quad ,
\end{eqnarray}
and the \sdss\ form of the de~Vaucouleurs (luv) profile is
\begin{eqnarray}\displaystyle
Q^{\luv}(\spos) &\equiv& \left\{\begin{array}{ll}
  \exp(-\alpha^{\luv}\,\left[[|\spos|^2 + 0.0004]^{1/8} - 1\right]) & \mbox{for~}|\spos| < 7 \\
  \exp(-\alpha^{\luv}\,\left[[|\spos|^2 + 0.0004]^{1/8} - 1\right])
  \,\left[1 - [|\spos| - 7]^2\right]^2 & \mbox{for~}7 < |\spos| < 8 \\
  0                                   & \mbox{for~}8 < |\spos|
\end{array}\right.
\\
\alpha^{\luv} &\equiv& 7.66925
\quad .
\end{eqnarray}
The half-light inverse-radius parameters $\alpha^{\exp}$ and
$\alpha^{\dev}$---and the softening and cutoff radius parameters---are
taken from the \sdss\ codebase.

The profiles above are normalized to have unit intensity
(approximately) at their half-light radii.  In many cases, the
investigator wants profiles that are normalized to have unit total
flux (intensity integrated over solid angle).  Although there is an
analytic result for the dev profile, numerical integration of the
concentrated profiles dev and luv to determine total fluxes can be
challenging.  This is not true for the MoG approximations: Each
Gaussian is normalized, so the sum of the amplitudes $\sum_m
a^{\luv}_m$ (for the luv profile, say) gives the total flux for the
MoG approximation to that profile.

We seek the \emph{best} MoG approximations.  This necessitates
definition of the word ``best''.  If we think of the profiles as being
two-dimensional probability distribution functions (for, say, the
arrivals of photons), then one natural choice is the K-L divergence or
similar cross-entropy or information-theoretic measure.  However, in
typical astronomical imaging, the galaxy is superimposed on a
substantial, flat sky level, and the noise in the data is close to
Gaussian.  This suggests more chi-squared-like objectives.  We adopt
the latter, in part because they are most appropriate for our specific
proposed application (modeling \sdss-like astronomical imaging), but
experiments we have performed suggest that information-theoretic
objectives also lead to good results.

In detail, the chi-squared objective we minimize---the
\emph{badness}---is a squared residual between the exact profile
function $Q(\spos)$ and its MoG approximation.  It is designed to be
equivalent to a chi-squared statistic in a homoskedastic
two-dimensional image of the profile taken with extremely high angular
resolution (pixels of size 0.001 the half-light radius) and vanishing
point-spread function.  Quantitatively the badness is defined to be
the mean squared residual in the $Q(\spos)$ functions, which are
normalized to have unit intensity at the half-light radius, averaged
over a two-dimensional circular region in the $\spos$ plane centered
on the (circularly symmetric) profile and extending out to radius
$\xi_{\max}$.  We use $\xi_{\max}=8$ for all profiles except the
$\lux$ profile, for which we use $\xi_{\max}=4$.  In practice, the
badness is computed in a one-dimensional numerical integral but the
integral is weighted in radius (weight increasing linearly with
radius) to make it equivalent to the two-dimensional chi-squared.  We
also add to the badness a very tiny coefficient (on the order of
$10^{-3}$ of the best-fit badness) times the sum of the variances
$v_m$ for regularization.  In practice, this term doesn't have much
effect and could be dropped.

Optimization (minimization) of the badness is performed by the
\project{scipy} implementation of the BFGS algorithm, with many
initializations to explore multiple local minima.  Further details are
available in the code, which is publicly
available.\footnote{\url{https://github.com/davidwhogg/TheTractor/}}

\begin{figure}
\includegraphics[width=\figurewidth]{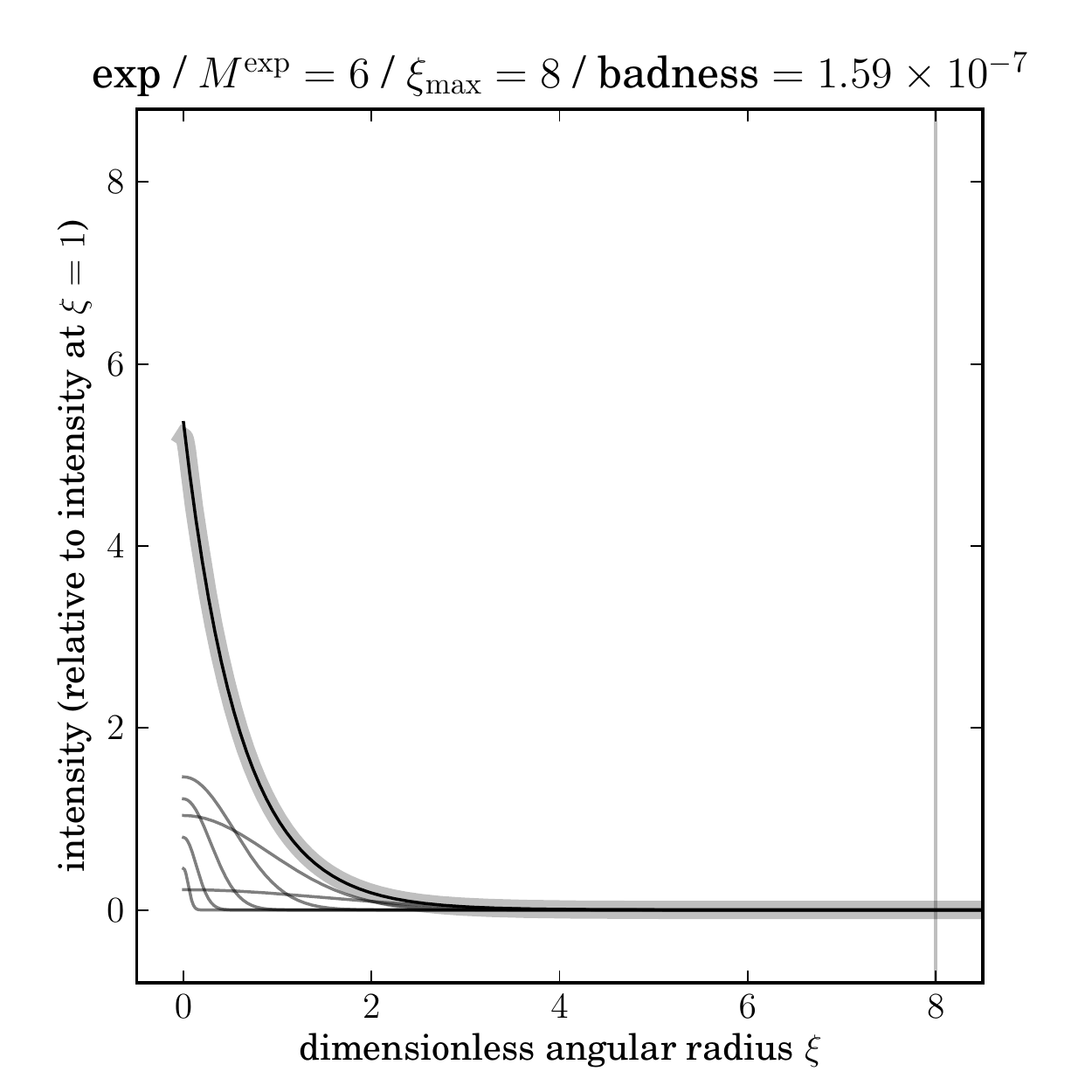}%
\includegraphics[width=\figurewidth]{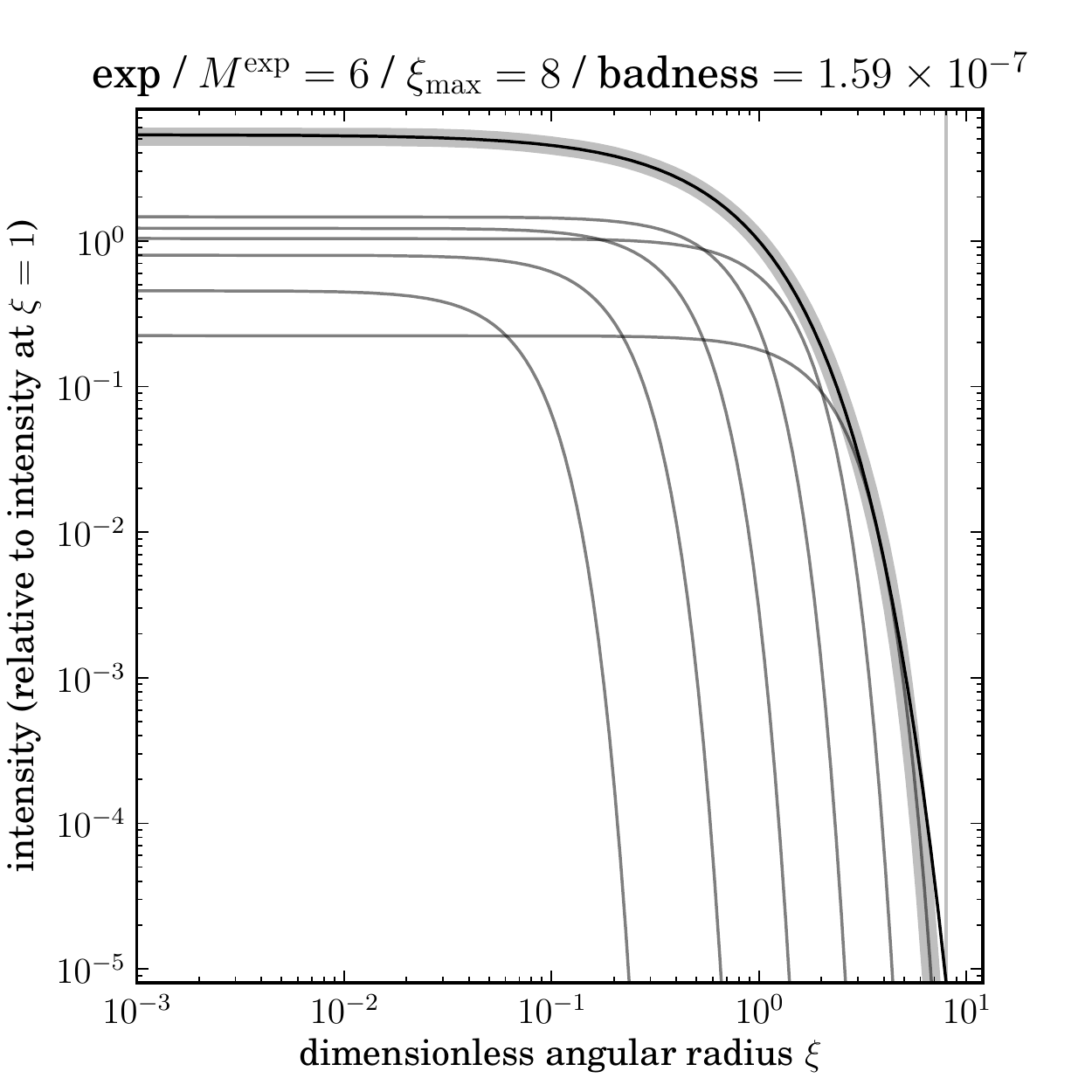}\\
\includegraphics[width=\figurewidth]{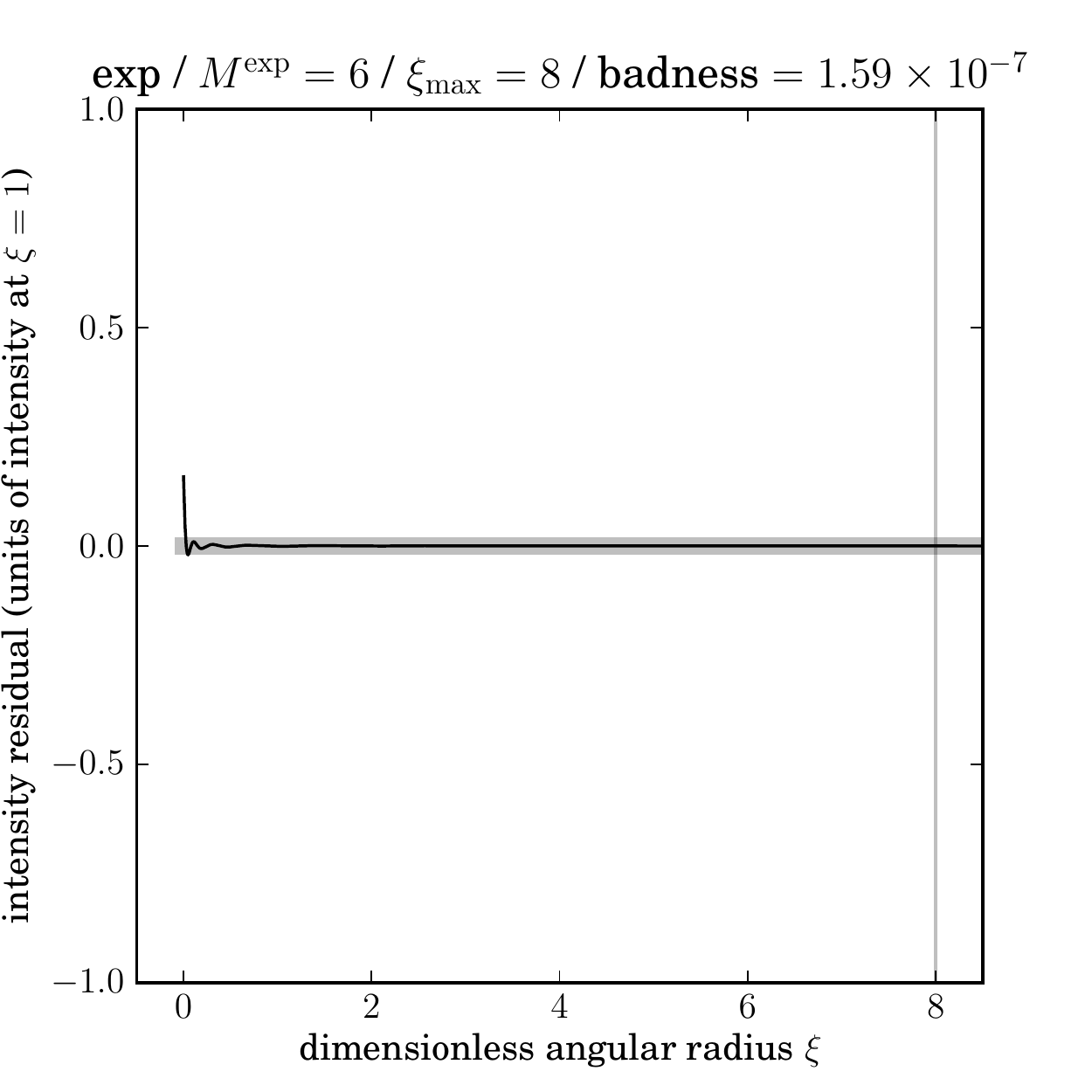}%
\includegraphics[width=\figurewidth]{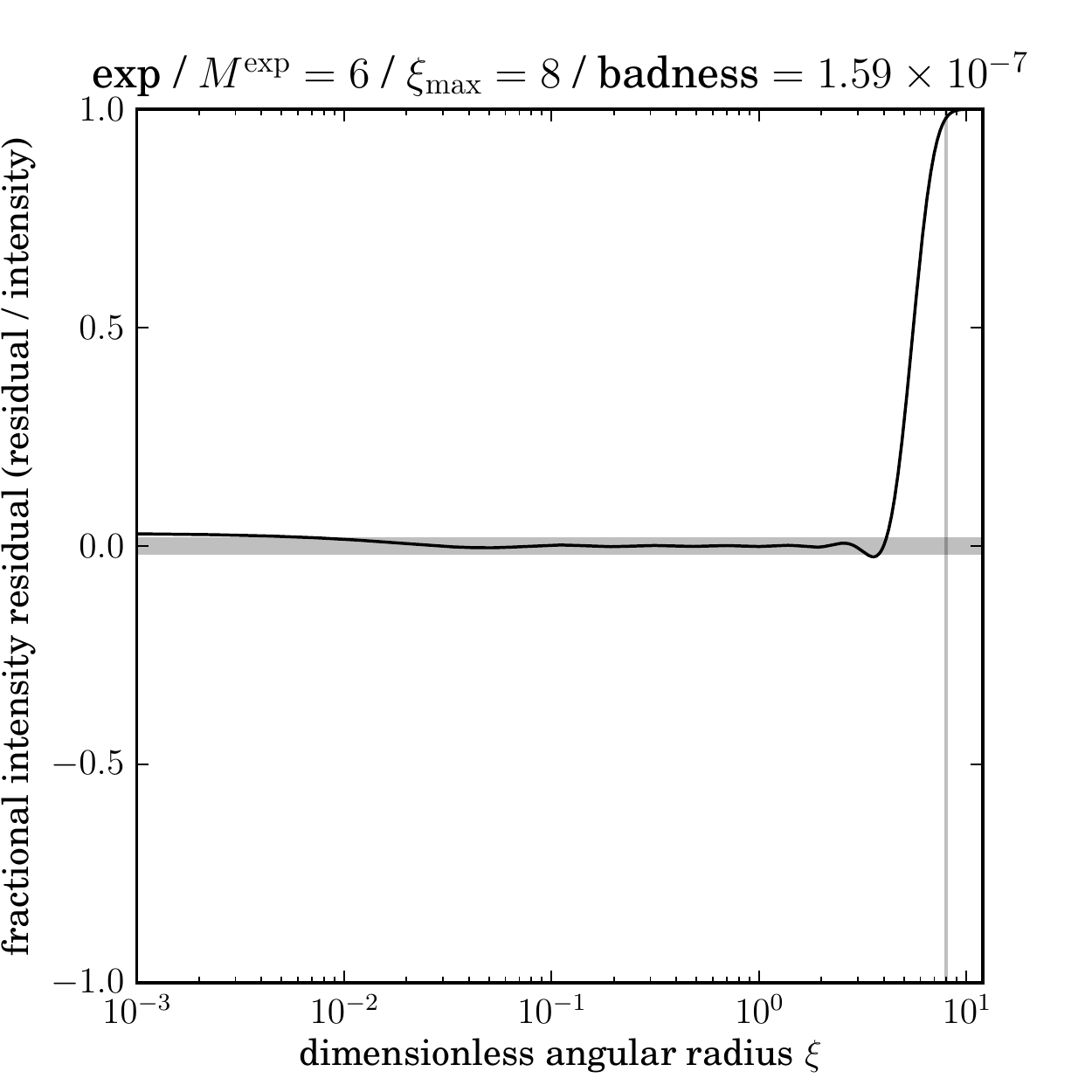}
\caption{\textsl{top-left:} The true exp profile (thin black line),
  the best $M^{\exp}=6$ mixture-of-Gaussian approximation (thick grey
  line), and the component Gaussians (multiplied by their
  corresponding amplitudes) contributing to the approximation (thin
  grey lines).  The plot title text gives $\xi_{\max}$ and the
  badness. \textsl{top-right:} The same but shown logarithmically.
  \textsl{bottom-left:} A representation of the residual or devation,
  on which the badness is computed.  \textsl{bottom-right:} The same
  but shown fractionally and logarithmically.\label{fig:exp}}
\end{figure}

\begin{figure}
\includegraphics[width=\figurewidth]{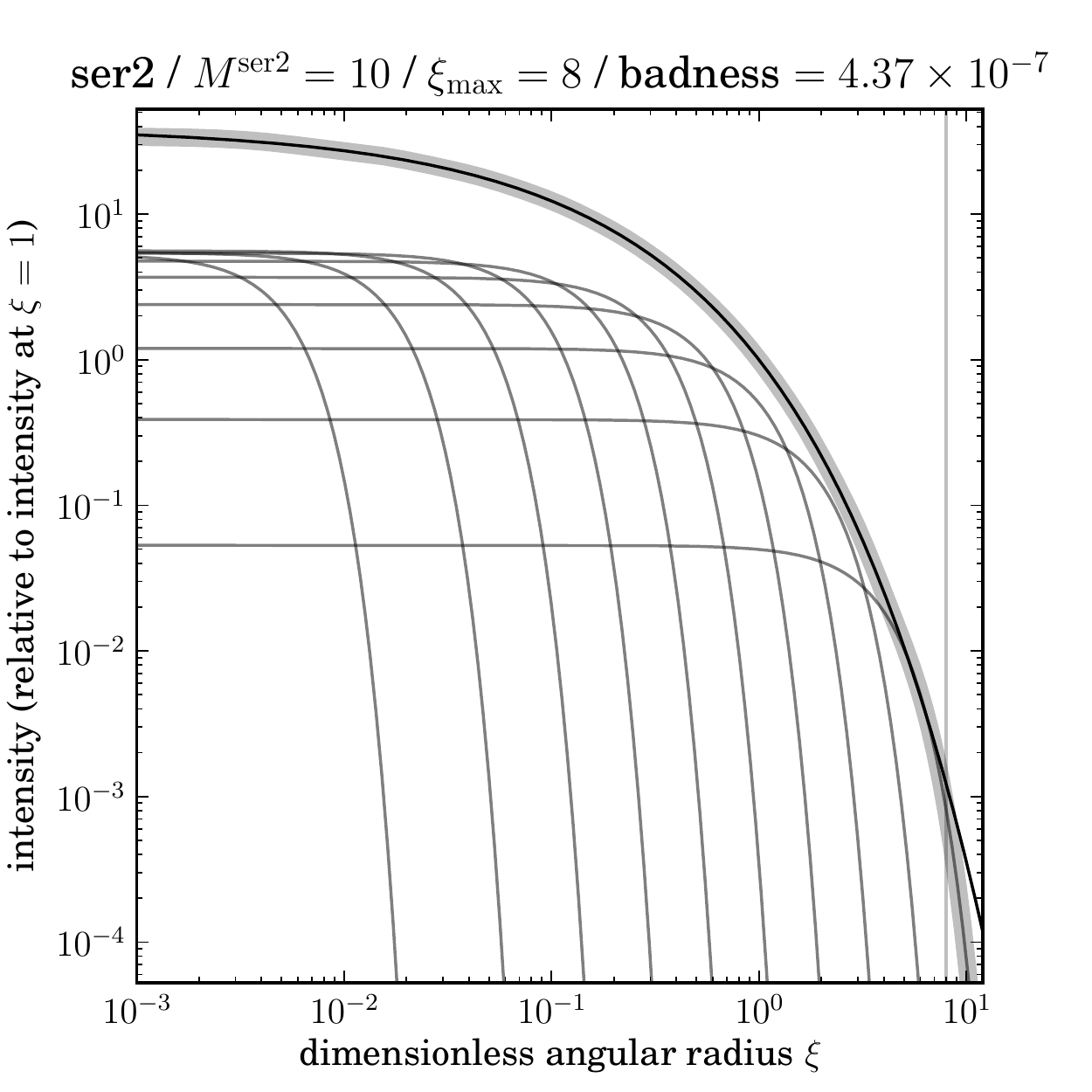}%
\includegraphics[width=\figurewidth]{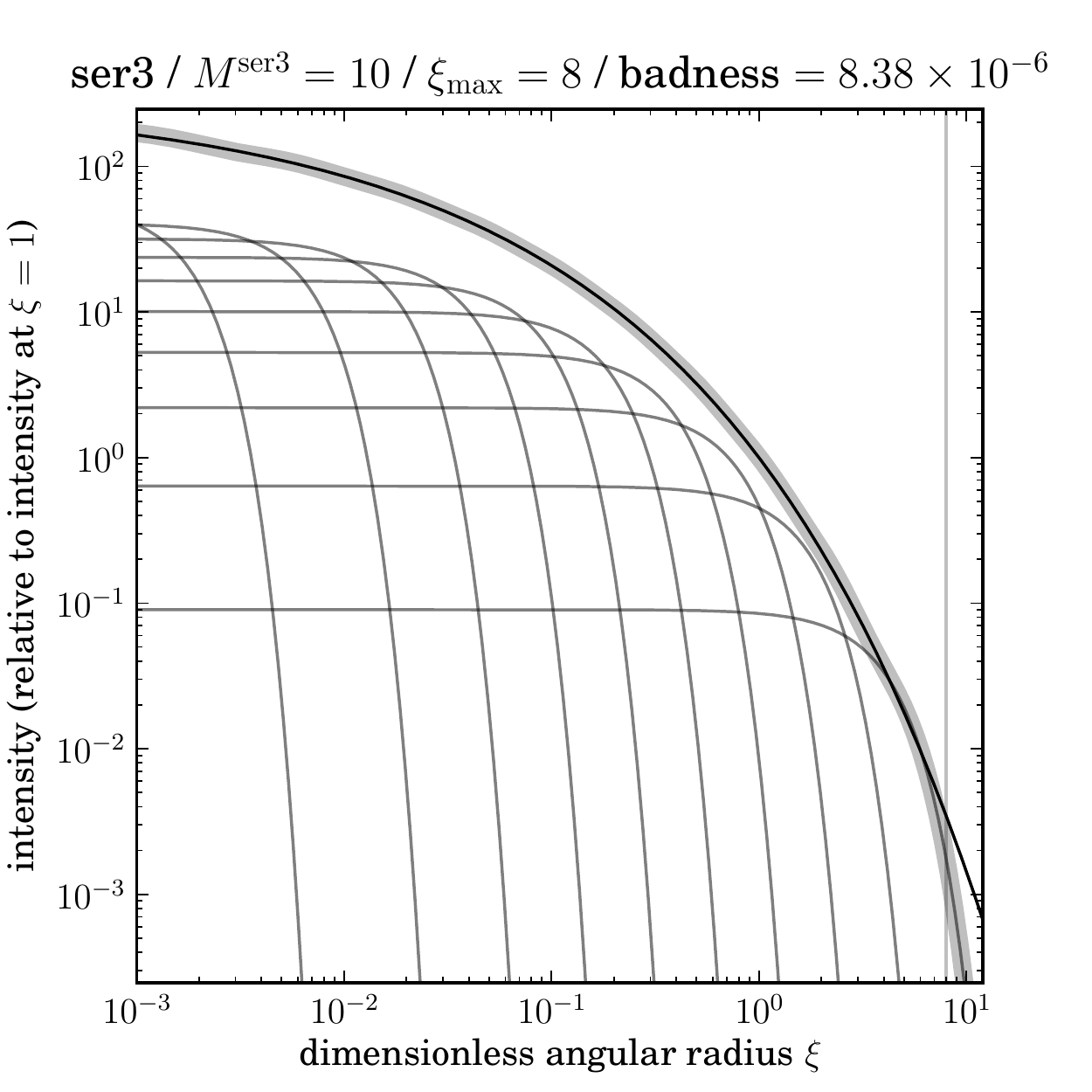}\\
\includegraphics[width=\figurewidth]{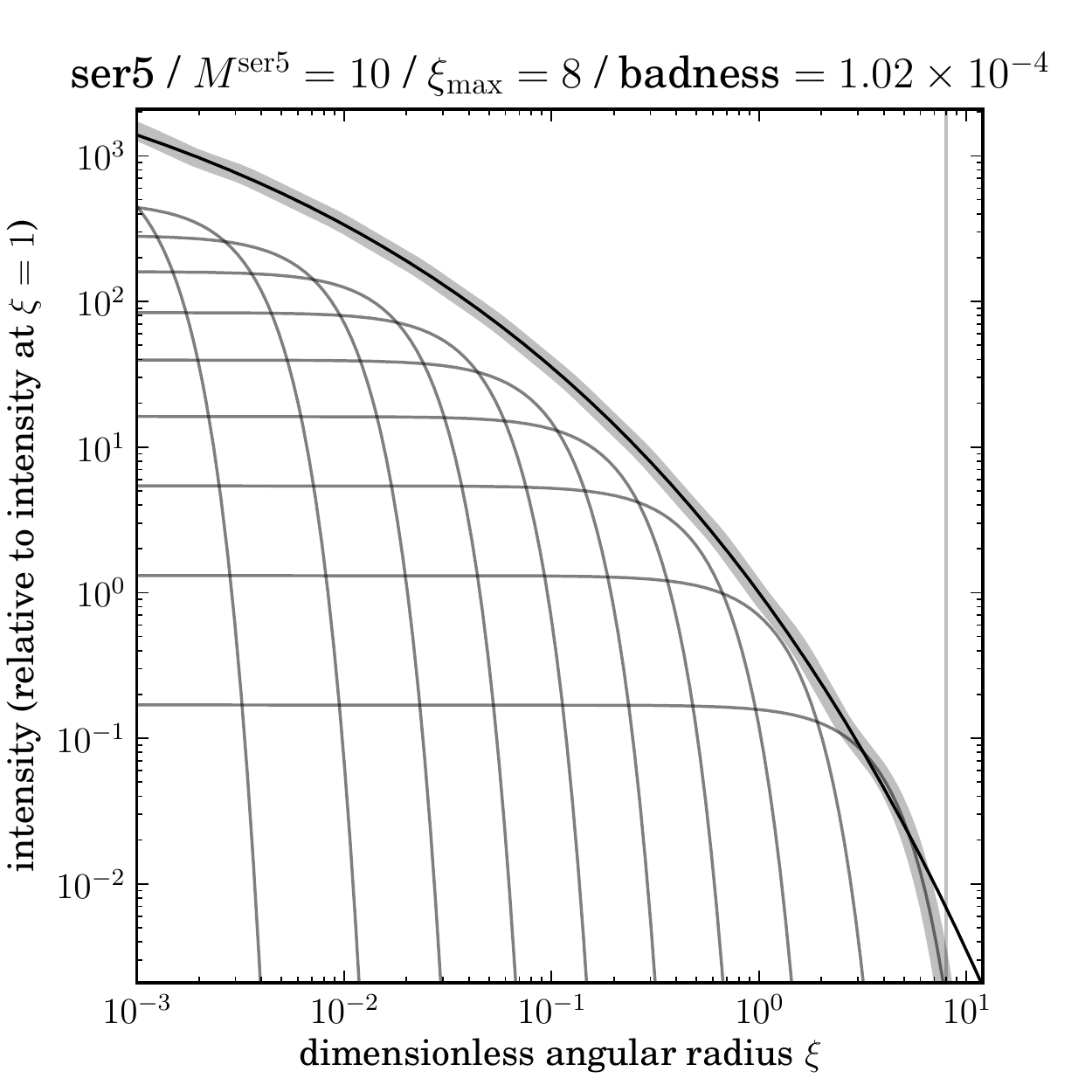}%
\includegraphics[width=\figurewidth]{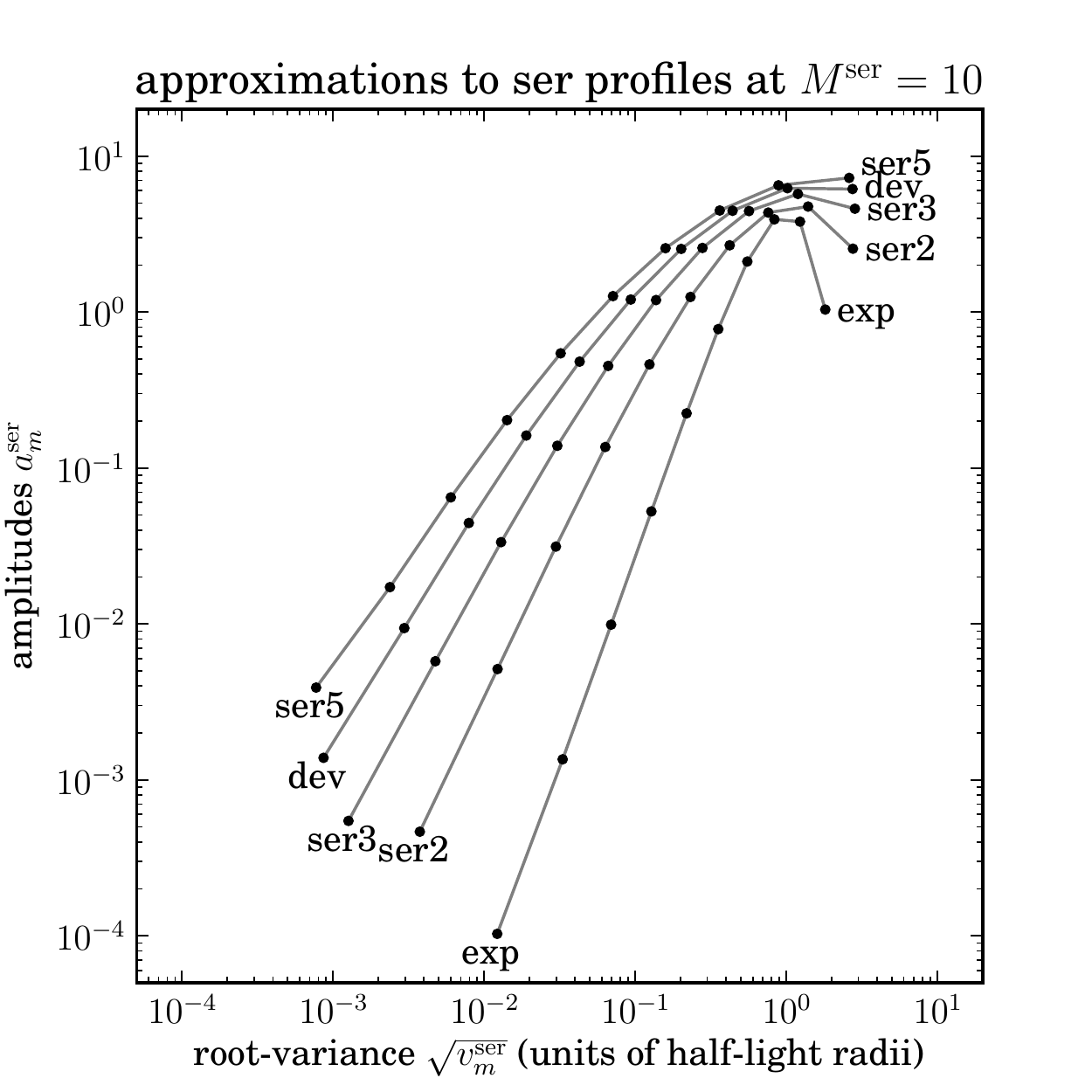}
\caption{Three ser profiles---with $n=2$, $3$, and $5$---and
  approximations.  The top-left, top-right, and bottom-left panels are
  equivalent to those in the top-right of \figurename~\ref{fig:exp}.
  \textsl{bottom-right:} The dependence on the amplitudes $a^{\ser(n)}_m$
  and root-variances $\sqrt{v^{\ser(n)}_m}$ on ser index $n$.\label{fig:ser}}
\end{figure}

\begin{figure}
\includegraphics[width=\figurewidth]{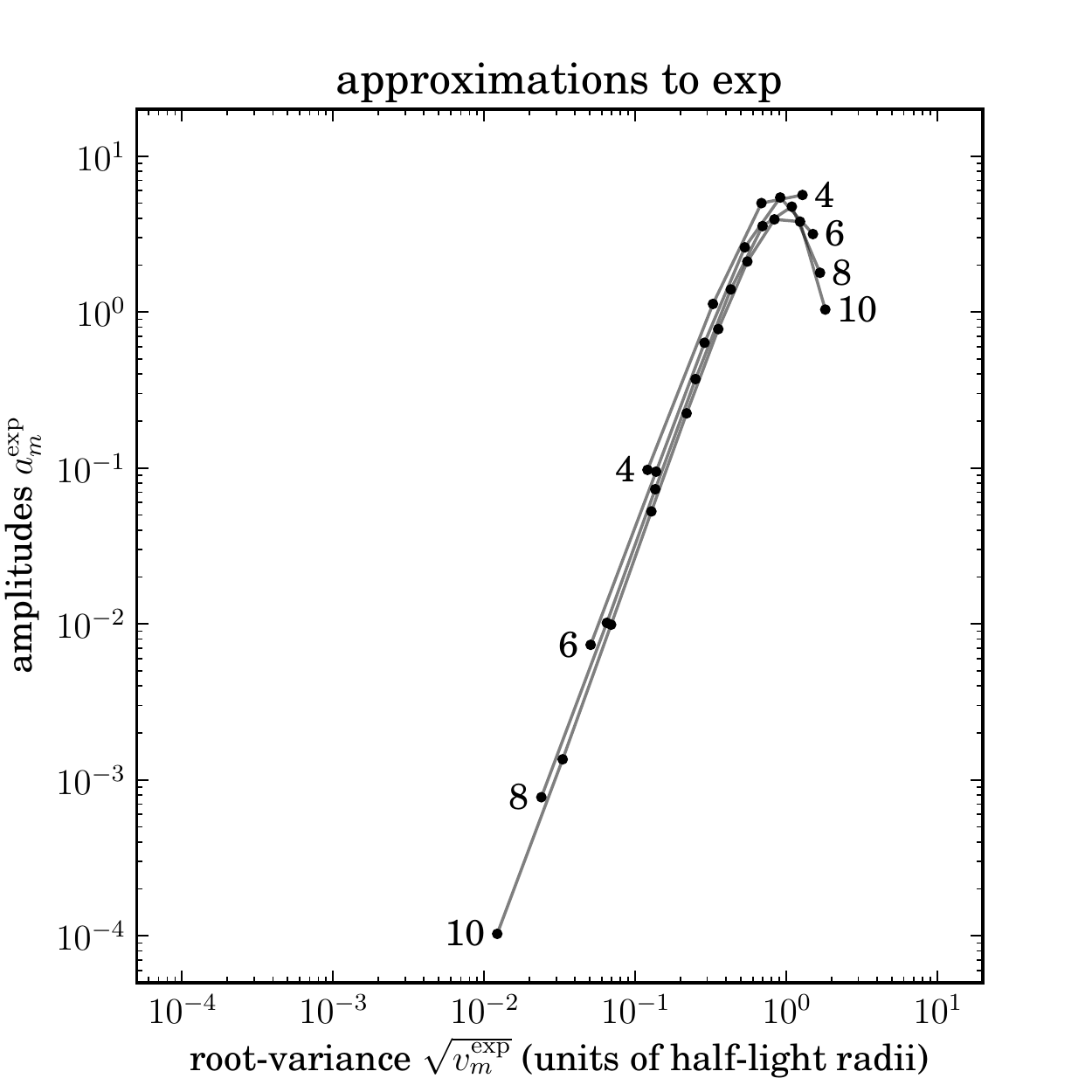}%
\includegraphics[width=\figurewidth]{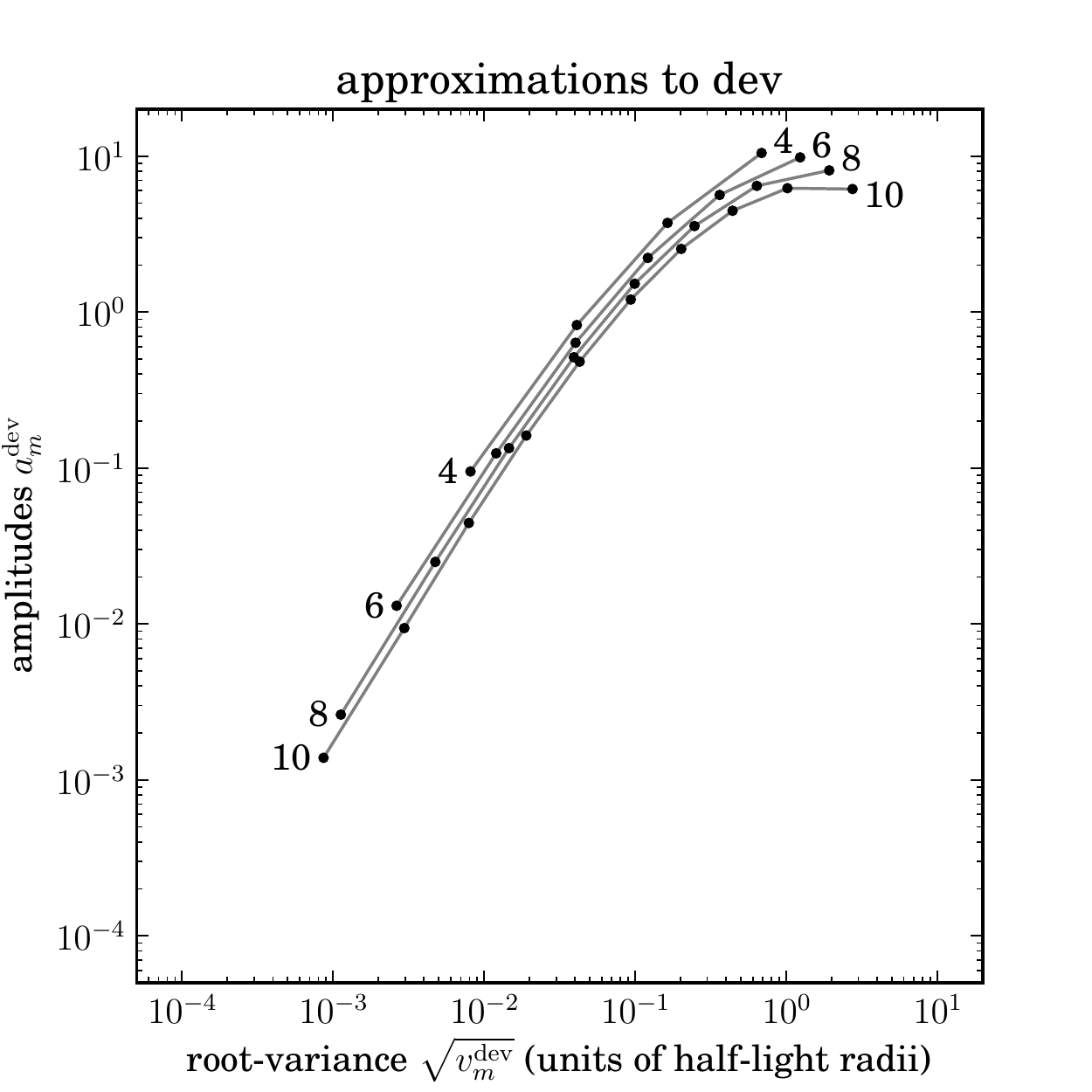}\\
\includegraphics[width=\figurewidth]{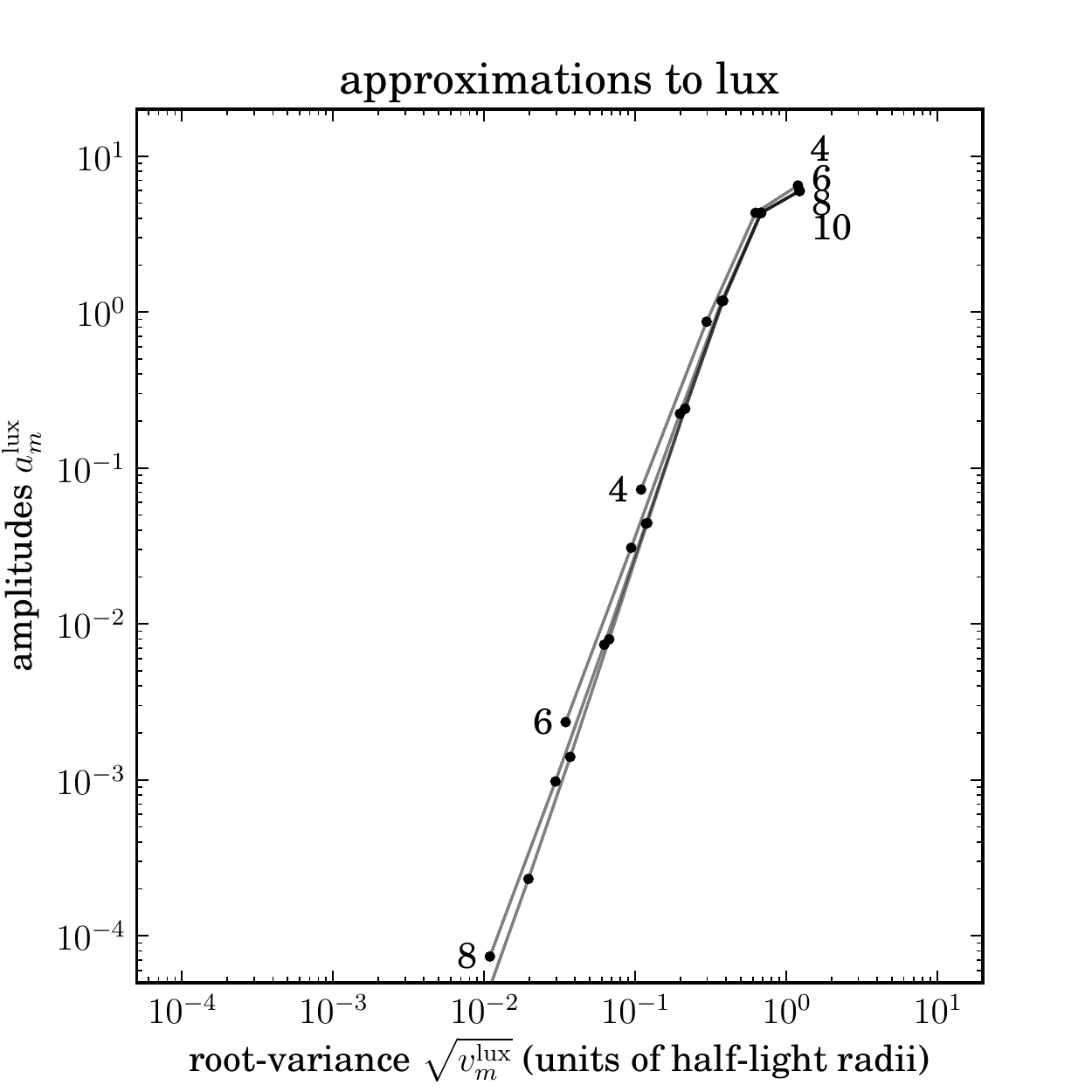}%
\includegraphics[width=\figurewidth]{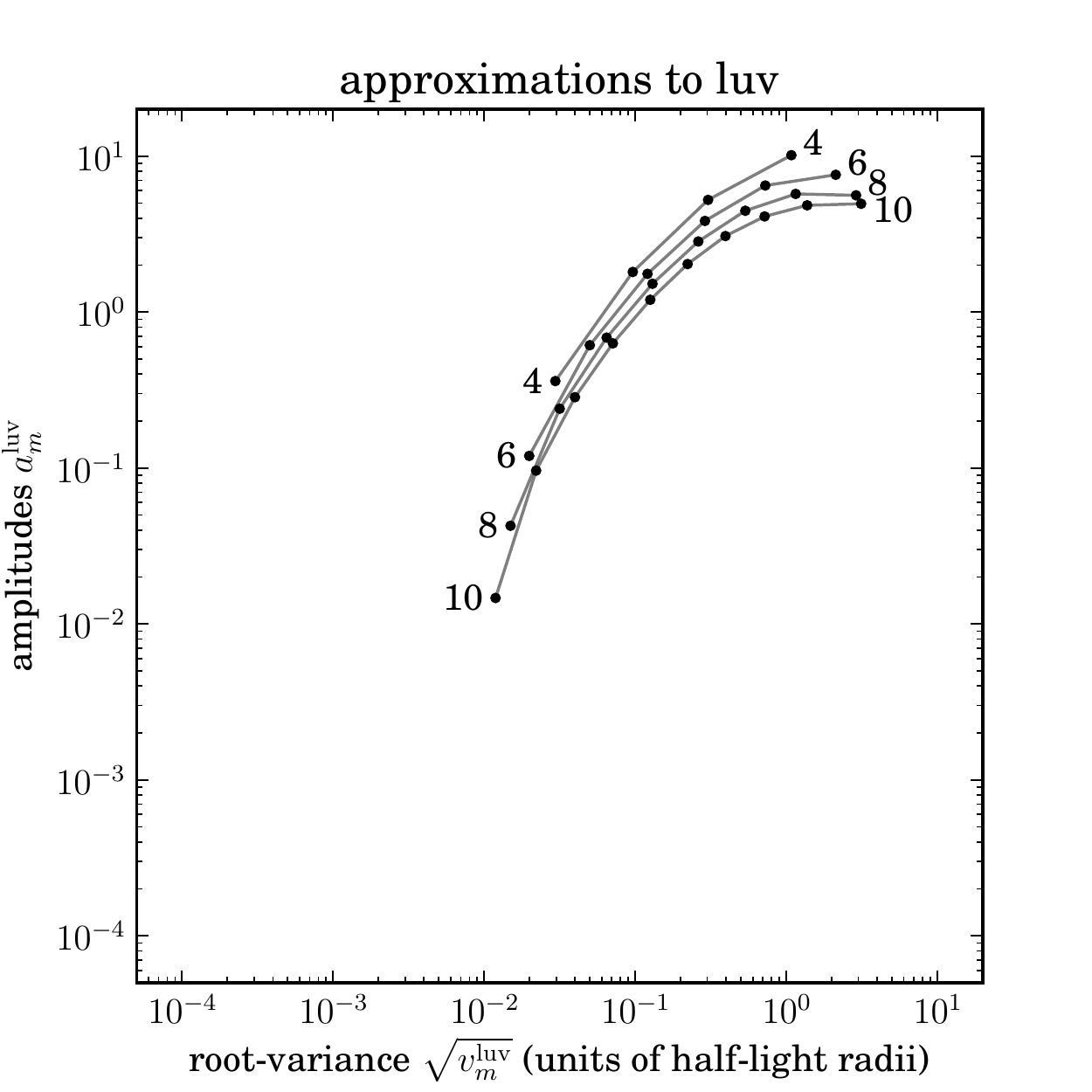}
\caption{Comparisons of approximations.  \textsl{top-left:} The
  dependence of amplitude $a^{\exp}_m$ and root-variance
  $\sqrt{v^{\exp}_m}$ on $M^{\exp}$ for the exp profile.
  \textsl{top-right:} The same but for the dev profile.
  \textsl{bottom-left:} The same but for the lux profile.
  \textsl{bottom-right:} The same but for the luv profile.\label{fig:M}}
\end{figure}

The results of the optimizations are shown in \tablename
s~\ref{tab:exp} and \ref{tab:lux} and \figurename s~\ref{fig:exp}
through \ref{fig:M}.  All the results shown in these figures and
tables and more are available in machine-readable form from author DWH
upon request.  In the \tablename s and \figurename s we show
root-variances rather than variances because these have units of
half-light radii; they are simple standard deviations for the Gaussian
components.

In our work on \project{The Tractor}, we use the $M^{\lux}=6$ lux and
the $M^{\luv}=8$ luv profiles.  Our advice to users would be to do the
same.  We use the lux and luv over the exp and dev partly because of
their better behaviors numerically, and partly because they and we are
both part of the \project{SDSS} tradition.  These---$M^{\lux}=6$ and
$M^{\luv}=8$---are good compromises between mixture complexity ($M$)
and quality of fit (badness).  Also, even the best-fitting late-type
and early-type galaxies deviate from exponential and de~Vaucouleurs
fits by more than do these high-quality MoG approximations; no
precision is lost.

In \figurename~\ref{fig:ser}, we show the dependence of amplitudes
$a^{\ser(n)}_m$ and variances $v^{\ser(n)}_m$ on the ser index $n$.
There is clearly continuity; a valuable follow-up project would be to
give expressions for the amplitudes and variances as a function of ser
index $n$.  In the absence of cleverness our advice would be to make
use of smooth interpolation.

The value of these MoG approximations comes when they are to be
convolved with a PSF (usually in fact a pixel-convolved PSF) that is
itself also represented as a MoG.  In this scenario, the PSF
$\psi(\Delta\pos)$---which is thought of as a function of focal-plane
displacement $\Delta\pos$ away from, say, a true stellar position---is
represented as a $K$-Gaussian MoG
\begin{eqnarray}\displaystyle
\psi(\Delta\pos) &=& \sum_{k=1}^K p_k\,\normal(\Delta\pos\given\mean_k,\var_k)
\\
1 &=& \sum_{k=1}^K p_k
\quad ,
\end{eqnarray}
where the means $\mean_k$ are \emph{not} required to vanish because
the PSF can have arbitrarily non-trivial structure (think speckles and
the like) and the variances $\var_k$ will not in general be
proportional to the identity or even diagonal because the PSF will not
in general be round.  An example that illustrates the use of this PSF
is the following: A star of flux $S_s$ at focal-plane position
$\pos_s$ will lead to an image (PSF-convolved intensity map) of the
form
\begin{eqnarray}\displaystyle
I(\pos\given\star,S_s,\pos_s) &=& \sum_{k=1}^K S_s\,p_k\,\normal(\pos\given\pos_s+\mean_k,\var_k)
\quad .
\end{eqnarray}
That is, when the PSF is represented as a MoG, any image of a
star---or indeed any image of any set of stars---is also represented
as a MoG.

\begin{figure}
\setlength{\figurewidth}{0.25\textwidth}
\begin{tabular}{@{}c@{}c@{}}
\includegraphics[width=\figurewidth]{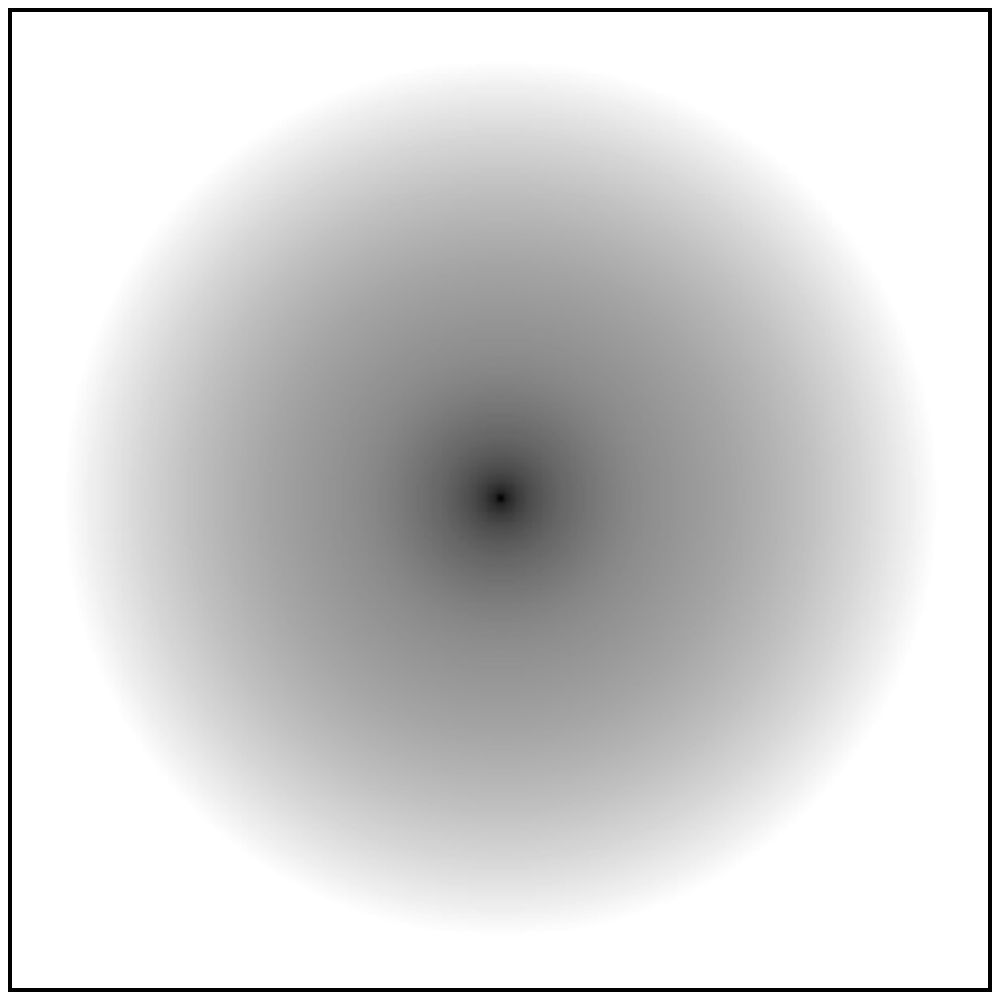} &
\includegraphics[width=\figurewidth]{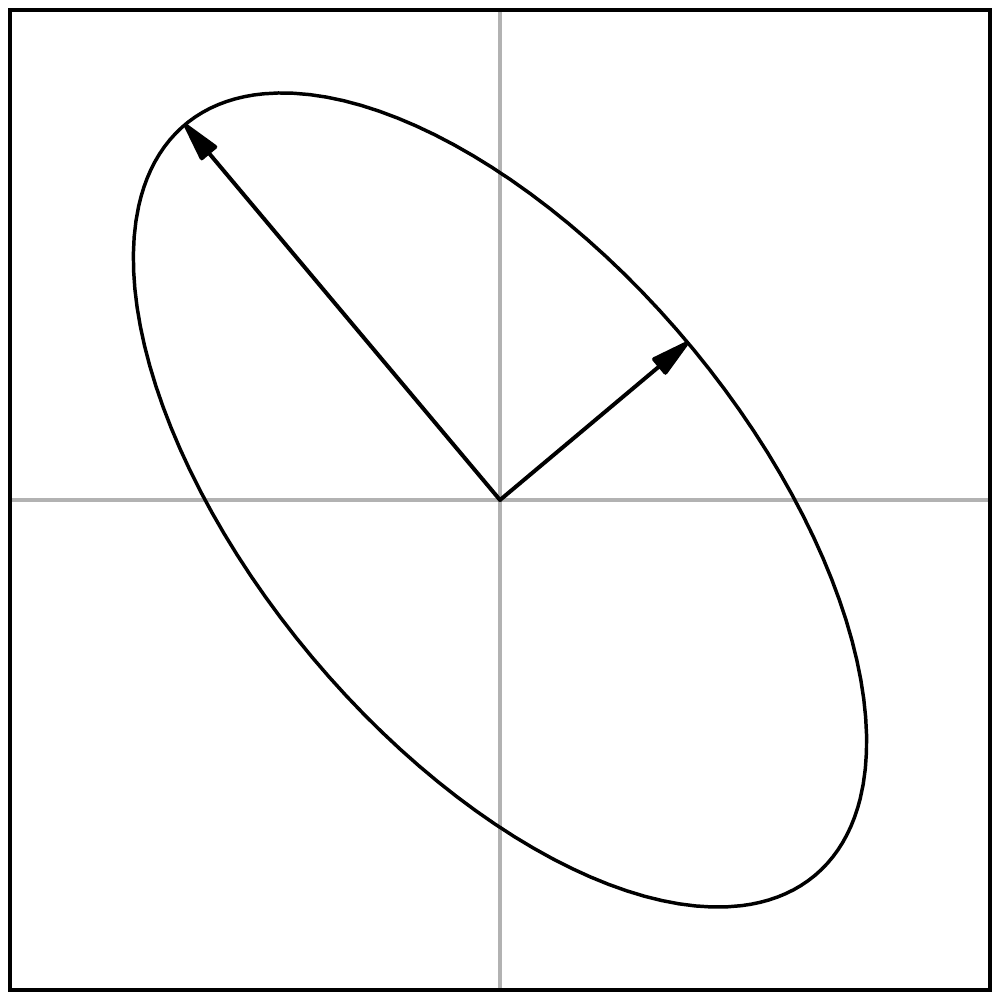} \\
\includegraphics[width=\figurewidth]{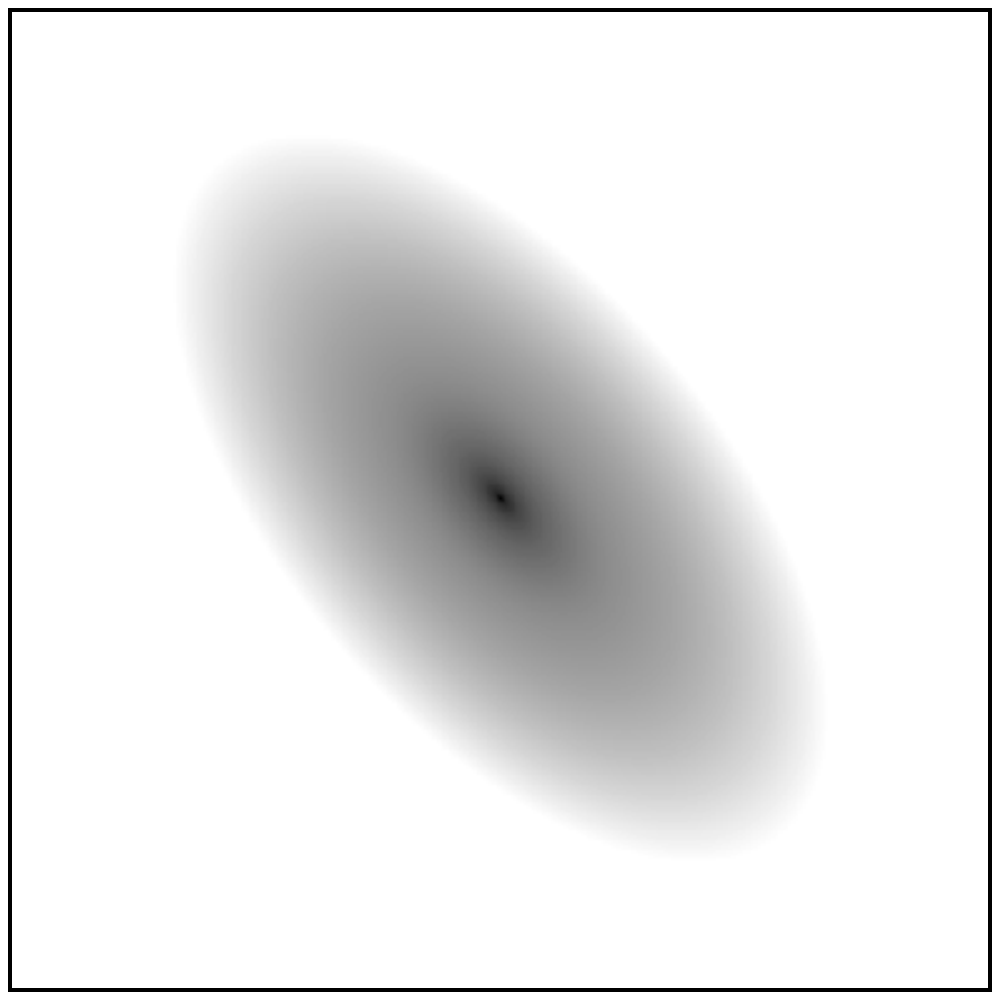} &
\includegraphics[width=\figurewidth]{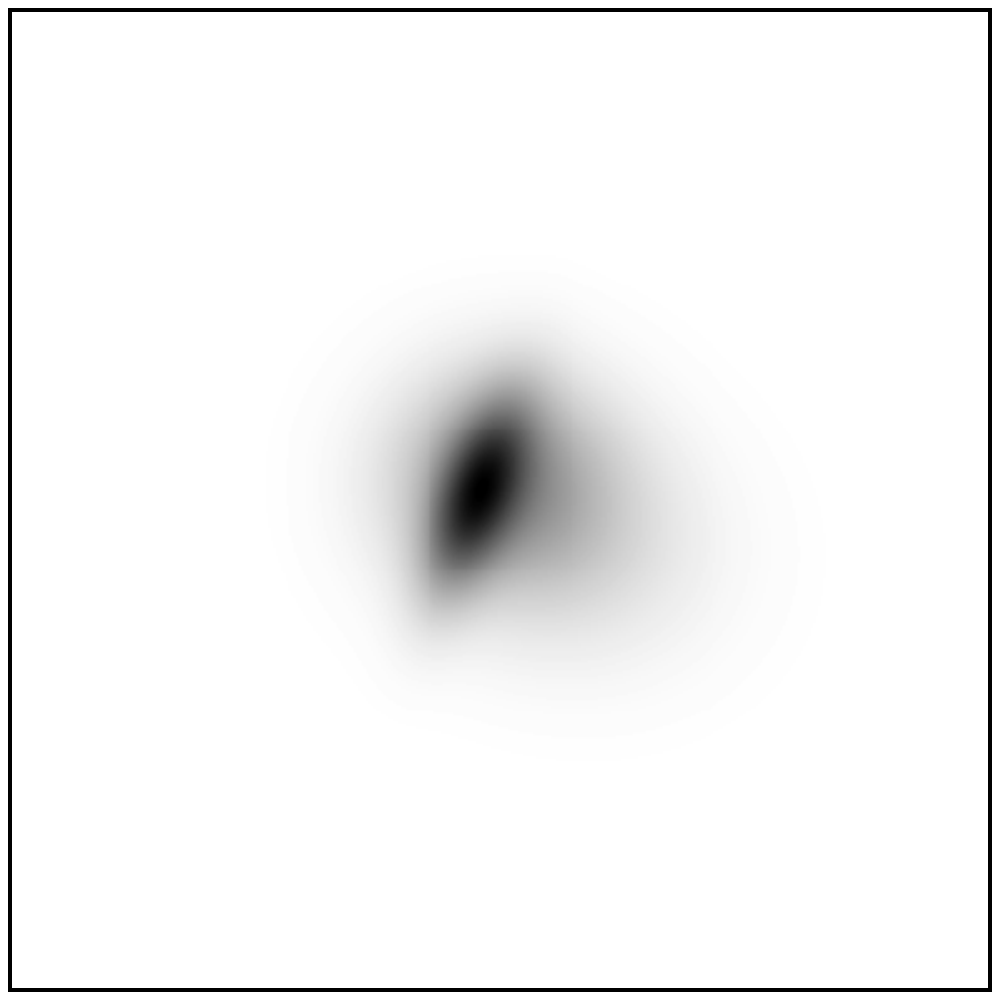} \\
\includegraphics[width=\figurewidth]{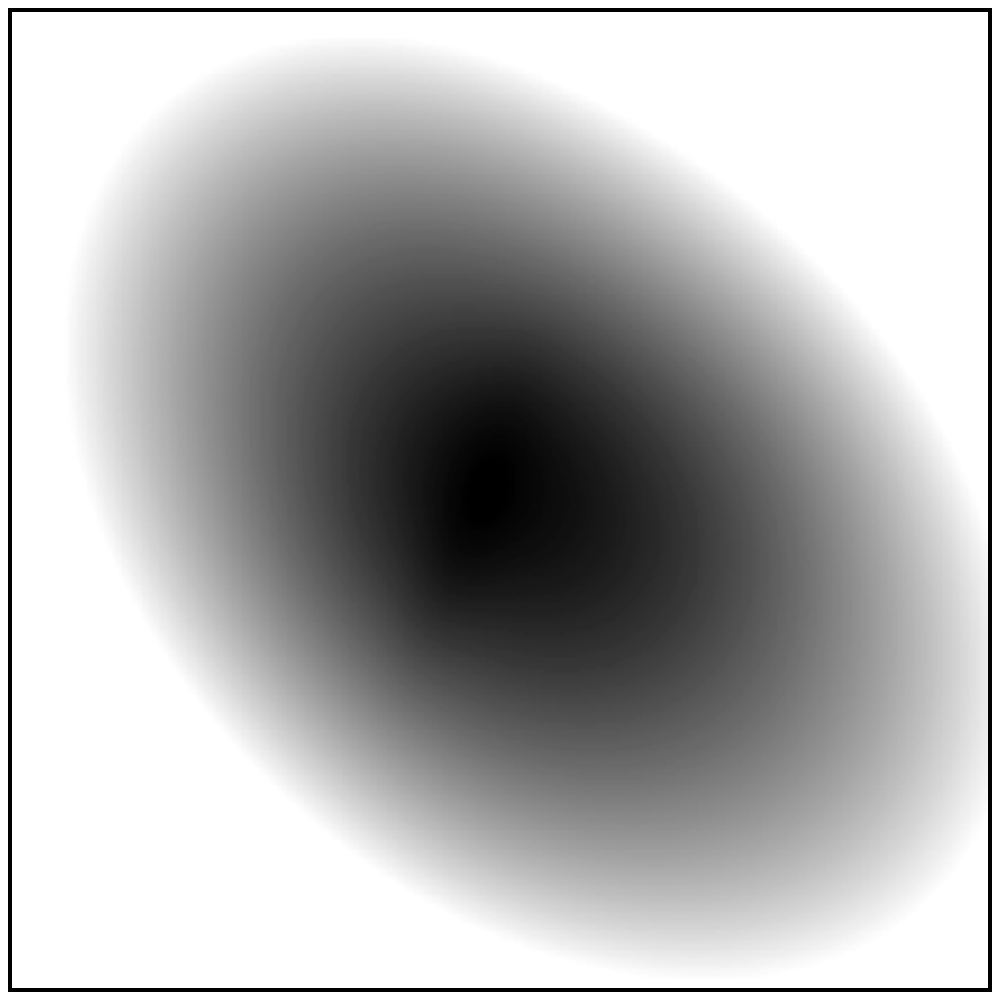} &
\includegraphics[width=\figurewidth]{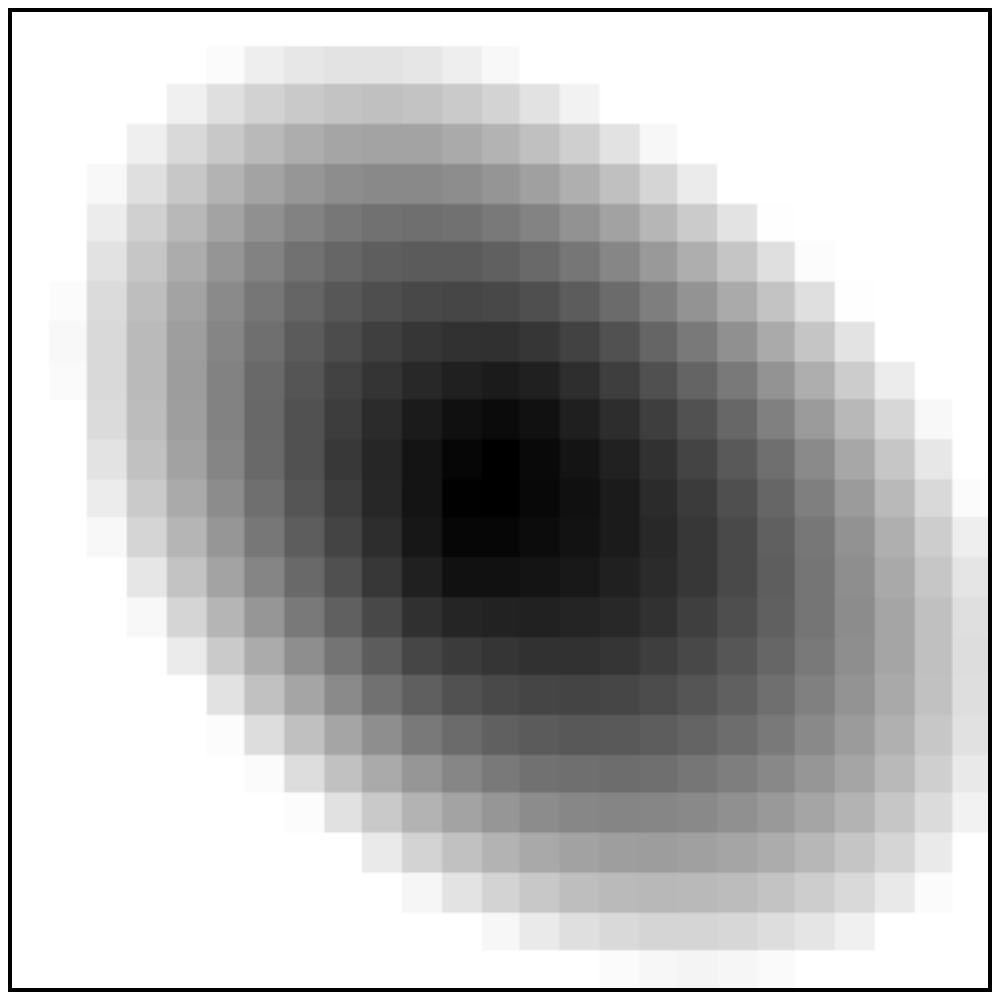}
\end{tabular}
\caption{Demonstration of use of the profiles, or the implicit
  generative model in this \documentname.  \textsl{top-left:} The
  circular dimensionless $M^{\luv}=8$ mixture-of-Gaussian
  approximation to the luv profile, represented on a very fine pixel
  grid. \textsl{top-right:} The ellipse representing the non-trivial
  affine transformation to be applied to the circular, dimensionless
  profile.  \textsl{middle-left:} The sheared profile.
  \textsl{middle-right:} A $K=3$ mixture-of-Gaussian model of the
  pixel-convolved point-spread function, represented on the very fine
  pixel grid.  \textsl{bottom-left:} The sheared profile convolved
  with the PSF, represented on the very fine pixel grid.
  \textsl{bottom-right:} The sheared luv convolved with the PSF, but
  now shown on a realistic pixel grid.  Because by assumption the PSF
  is a pixel-convolved PSF, the representation on the realistic grid
  is found simply by interpolating to the pixel centers the
  mixture-component Gaussians.\label{fig:example}}
\end{figure}

Applying this PSF to an exp or dev galaxy is slightly more
complicated, because the galaxy has not just a flux $S_g$ and a
central position $\pos_g$; it also has a shape.  Because we are only
considering these simple galaxies, we are only permitting ellipsoidal
shapes, which can be represented by a semi-major axis $a$, a
semi-minor axis $b$, and a position angle $\phi$, or equivalently by
eigenvalues $a, b$ and eigenvectors $\uv_1, \uv_2$, or equivalently by
an affine transformation $\affine_g$ that takes a circle to the
relevant ellipse (and is therefore a general representation of an
ellipse; it is also the matrix square root of the symmetric variance
tensor describing the ellipse).  The galaxy is distorted by this
affine transformation \emph{prior} to PSF convolution, so the
focal-plane image (PSF-convolved intensity field) for a general (say)
$\exp$ galaxy is given by
\begin{eqnarray}\displaystyle
I(\pos\given\exp,S_g,\pos_g,\affine_g) &=& \sum_{k=1}^K \sum_{m=1}^{M^{\exp}} S_g\,a^{\exp}_m\,p_k\,\normal(\pos\given\pos_g+\mean_k,\var_{gm}+\var_k)
\\
\var_{gm} &\equiv& \affine_g\cdot\var^{\exp}_m\cdot\transpose{\affine_g}
\\
\affine_g &=& \left[a\,\uv_1 , b\,\uv_2 \right]
\quad ,
\end{eqnarray}
where $a$ and $b$ are the major and minor axis lengths of the galaxy
ellipse (in appropriate units) and $\uv_1$ and $\uv_2$ are the
eigenvectors in image coordinates pointing in the major-axis and
minor-axis directions respectively.  Implicitly all vectors are
two-dimensional column vectors, and $\affine_g$ is a $2\times 2$
affine transformation matrix that contains the ``shape'' (position
angle, major-axis, and ellipticity) information about the galaxy.  The
$\dev$, $\ser(n)$, $\luv$, and $\lux$ cases are all essentially the
same.  Note the important and key result of this \documentname, to
wit, that a MoG galaxy model (with $M$ components) convolved with a
MoG PSF model (with $K$ components) yields a MoG model image (with
$[M\,K]$ components).  In \figurename~\ref{fig:example} we show how we
are using these MoG approximations in \project{The Tractor}---a
generative modeling framework for measuring astronomical objects---to
render PSF-convolved galaxy images.

In the above we said ``pixel-convolved PSF''.  In every context, when
modeling images, it is valuable to use the pixel-convolved PSF.  With
this definition of the PSF, the pixelized image is the PSF-convolved
true model evaluated at the pixel centers.  This operation is fast.
Other definitions for the PSF (the non-pixel-convolved, for example)
require that the user do \emph{two} convolutions, the first with the
PSF and the second with the square (or worse) pixel.  Our advice: Only
fit for and use pixel-convolved PSFs.

If your PSF is not in MoG form, it is still the case that convolution
of a MoG approximation of a dev (say) profile will in general be
easier than convolution of the original dev profile.  The reason is
that convolution of a Gaussian with any PSF is fast (indeed most
image-processing languages have such functions built in); the
PSF-convolved profile becomes in this case just a mixture of
Gaussian-convolved PSFs.

The speed-ups that can be obtained by using MoG approximations can be
very large.  In our image-modeling project \project{The Tractor}, we
were PSF-convolving by rendering the profiles (especially the profile
centers) at very high resolution (hundreds to thousands of resolution
elements in the central pixel are necessary for good precision on the
dev profile).  We were then convolving that high-resolution model with
a low-resolution PSF and rendering to a low-resolution image pixel
grid.  These expensive operations were obviated by the MoG profiles,
which involve only rendering a small number of Gaussians at the pixel
centers on the low-resolution pixel grid.  The MoG approximations
saved us more than an order of magnitude in compute time, especially
in optimization, where derivatives have to be taken with respect to
the unconvolved model properties.

In the \project{SDSS}, \project{GALFIT}, and much of our own work, the
models that are fit are (effectively) mixtures of exp and dev or exp
and ser or lux and luv profiles.  Mixtures of profiles that are each
themselves mixtures of Gaussians are no harder to render than either
profile separately.  There is some book-keeping, of course, because
each component gets affine-transformed separately before they are both
PSF-convolved.

One amusing aspect of MoG profiles has to do with projection from
three to two dimensions.  The projection of a three-dimensional
Gaussian is a two-dimensional Gaussian; the two-dimensional, rigid,
circular approximations we have made for the ser profiles can be
deprojected to rigid, spherical approximations to the
three-dimensional profiles trivially.  The two-dimensional models we
have started with are not accurate models of galaxies in detail---no
galaxy follows exactly any ser profile---so deprojection of our
approximations are not that interesting in themselves.  However,
the general program of fitting two-dimensional sources with MoGs may
have strong implications in the future for three-dimensional modeling
and deprojection.

\acknowledgements It is a pleasure to thank
      Brendon Brewer (Auckland),
      Jim Bosch (Princeton),
  and Kevin Bundy (IPMU)
for valuable comments.  This work was supported in part by NASA (grant
NNX12AI50G) and the NSF (grant IIS-1124794).  All the results shown in
the figures and tables and more are available in machine-readable form
from author DWH upon request.

\clearpage
\begin{table}
\begin{tabular}{c|cc|cc|cc|}
&
\multicolumn{6}{|c|}{exp} \\
\hline
$M^{\exp}=$ &
 \multicolumn{2}{|c|}{$4$} &
 \multicolumn{2}{|c|}{$6$} &
 \multicolumn{2}{|c|}{$8$} \\
$m$ &
 $a^{\exp}_m$ & $\sqrt{v^{\exp}_m}$ &
 $a^{\exp}_m$ & $\sqrt{v^{\exp}_m}$ &
 $a^{\exp}_m$ & $\sqrt{v^{\exp}_m}$ \\
$1$ &
 $0.09733$ & $0.12068$ &
 $0.00735$ & $0.05072$ &
 $0.00077$ & $0.02394$ \\
$2$ &
 $1.12804$ & $0.32730$ &
 $0.09481$ & $0.13756$ &
 $0.01017$ & $0.06492$ \\
$3$ &
 $4.99846$ & $0.68542$ &
 $0.63572$ & $0.28781$ &
 $0.07313$ & $0.13581$ \\
$4$ &
 $5.63632$ & $1.28089$ &
 $2.60077$ & $0.53195$ &
 $0.37184$ & $0.25095$ \\
$5$ &
 $~$ & $~$ &
 $5.42848$ & $0.91209$ &
 $1.39736$ & $0.42942$ \\
$6$ &
 $~$ & $~$ &
 $3.16445$ & $1.50157$ &
 $3.56100$ & $0.69675$ \\
$7$ &
 $~$ & $~$ &
 $~$ & $~$ &
 $4.74338$ & $1.08885$ \\
$8$ &
 $~$ & $~$ &
 $~$ & $~$ &
 $1.78684$ & $1.67302$ \\
\hline
$\sum_m a^{\exp}_m=$ &
 \multicolumn{2}{|c|}{$11.860$} &
 \multicolumn{2}{|c|}{$11.932$} &
 \multicolumn{2}{|c|}{$11.944$} \\
badness~$=$ &
 \multicolumn{2}{|c|}{$4.35\times 10^{-6}$} &
 \multicolumn{2}{|c|}{$1.59\times 10^{-7}$} &
 \multicolumn{2}{|c|}{$8.90\times 10^{-9}$} \\
\end{tabular}
\\
\vspace{2ex}\\
\begin{tabular}{c|cc|cc|cc|}
&
\multicolumn{6}{|c|}{dev} \\
\hline
$M^{\dev}=$ &
 \multicolumn{2}{|c|}{$6$} &
 \multicolumn{2}{|c|}{$8$} &
 \multicolumn{2}{|c|}{$10$} \\
$m$ &
 $a^{\dev}_m$ & $\sqrt{v^{\dev}_m}$ &
 $a^{\dev}_m$ & $\sqrt{v^{\dev}_m}$ &
 $a^{\dev}_m$ & $\sqrt{v^{\dev}_m}$ \\
$1$ &
 $0.01308$ & $0.00263$ &
 $0.00262$ & $0.00113$ &
 $0.00139$ & $0.00087$ \\
$2$ &
 $0.12425$ & $0.01202$ &
 $0.02500$ & $0.00475$ &
 $0.00941$ & $0.00296$ \\
$3$ &
 $0.63551$ & $0.04031$ &
 $0.13413$ & $0.01462$ &
 $0.04441$ & $0.00792$ \\
$4$ &
 $2.22560$ & $0.12128$ &
 $0.51326$ & $0.03930$ &
 $0.16162$ & $0.01902$ \\
$5$ &
 $5.63989$ & $0.36229$ &
 $1.52005$ & $0.09926$ &
 $0.48121$ & $0.04289$ \\
$6$ &
 $9.81523$ & $1.23604$ &
 $3.56204$ & $0.24699$ &
 $1.20357$ & $0.09351$ \\
$7$ &
 $~$ & $~$ &
 $6.44845$ & $0.63883$ &
 $2.54182$ & $0.20168$ \\
$8$ &
 $~$ & $~$ &
 $8.10105$ & $1.92560$ &
 $4.46441$ & $0.44126$ \\
$9$ &
 $~$ & $~$ &
 $~$ & $~$ &
 $6.22820$ & $1.01833$ \\
$10$ &
 $~$ & $~$ &
 $~$ & $~$ &
 $6.15393$ & $2.74555$ \\
\hline
$\sum_m a^{\dev}_m=$ &
 \multicolumn{2}{|c|}{$18.454$} &
 \multicolumn{2}{|c|}{$20.307$} &
 \multicolumn{2}{|c|}{$21.290$} \\
badness~$=$ &
 \multicolumn{2}{|c|}{$2.01\times 10^{-3}$} &
 \multicolumn{2}{|c|}{$3.16\times 10^{-4}$} &
 \multicolumn{2}{|c|}{$3.90\times 10^{-5}$} \\
\end{tabular}

\caption{The amplitudes and root-variances for the best
  mixture-of-Gaussian approximations to the exp and dev profiles, for
  different mixture sizes.  The total (dimensionless) fluxes and the
  badnesses are given for each approximation.\label{tab:exp}}
\end{table}

\clearpage
\begin{table}
\begin{tabular}{c|cc|cc|cc|}
&
\multicolumn{6}{|c|}{lux} \\
\hline
$M^{\lux}=$ &
 \multicolumn{2}{|c|}{$4$} &
 \multicolumn{2}{|c|}{$6$} &
 \multicolumn{2}{|c|}{$8$} \\
$m$ &
 $a^{\lux}_m$ & $\sqrt{v^{\lux}_m}$ &
 $a^{\lux}_m$ & $\sqrt{v^{\lux}_m}$ &
 $a^{\lux}_m$ & $\sqrt{v^{\lux}_m}$ \\
$1$ &
 $0.07275$ & $0.10938$ &
 $0.00235$ & $0.03465$ &
 $0.00007$ & $0.01092$ \\
$2$ &
 $0.86763$ & $0.29694$ &
 $0.03080$ & $0.09405$ &
 $0.00098$ & $0.02966$ \\
$3$ &
 $4.33214$ & $0.62601$ &
 $0.22336$ & $0.19785$ &
 $0.00736$ & $0.06241$ \\
$4$ &
 $6.48325$ & $1.19571$ &
 $1.17949$ & $0.37413$ &
 $0.04404$ & $0.11794$ \\
$5$ &
 $~$ & $~$ &
 $4.33874$ & $0.67894$ &
 $0.24005$ & $0.21345$ \\
$6$ &
 $~$ & $~$ &
 $5.99821$ & $1.22540$ &
 $1.18175$ & $0.38155$ \\
$7$ &
 $~$ & $~$ &
 $~$ & $~$ &
 $4.31918$ & $0.68169$ \\
$8$ &
 $~$ & $~$ &
 $~$ & $~$ &
 $5.97985$ & $1.22635$ \\
\hline
$\sum_m a^{\lux}_m=$ &
 \multicolumn{2}{|c|}{$11.756$} &
 \multicolumn{2}{|c|}{$11.773$} &
 \multicolumn{2}{|c|}{$11.773$} \\
badness~$=$ &
 \multicolumn{2}{|c|}{$1.37\times 10^{-5}$} &
 \multicolumn{2}{|c|}{$4.64\times 10^{-6}$} &
 \multicolumn{2}{|c|}{$4.55\times 10^{-6}$} \\
\end{tabular}
\\
\vspace{2ex}\\
\begin{tabular}{c|cc|cc|cc|}
&
\multicolumn{6}{|c|}{luv} \\
\hline
$M^{\luv}=$ &
 \multicolumn{2}{|c|}{$6$} &
 \multicolumn{2}{|c|}{$8$} &
 \multicolumn{2}{|c|}{$10$} \\
$m$ &
 $a^{\luv}_m$ & $\sqrt{v^{\luv}_m}$ &
 $a^{\luv}_m$ & $\sqrt{v^{\luv}_m}$ &
 $a^{\luv}_m$ & $\sqrt{v^{\luv}_m}$ \\
$1$ &
 $0.11960$ & $0.01988$ &
 $0.04263$ & $0.01496$ &
 $0.01468$ & $0.01190$ \\
$2$ &
 $0.61327$ & $0.05008$ &
 $0.24013$ & $0.03166$ &
 $0.09627$ & $0.02210$ \\
$3$ &
 $1.75843$ & $0.12067$ &
 $0.68591$ & $0.06471$ &
 $0.28454$ & $0.03995$ \\
$4$ &
 $3.84242$ & $0.28955$ &
 $1.51937$ & $0.13017$ &
 $0.63005$ & $0.07117$ \\
$5$ &
 $6.48187$ & $0.72628$ &
 $2.83627$ & $0.26170$ &
 $1.19909$ & $0.12586$ \\
$6$ &
 $7.59437$ & $2.12717$ &
 $4.46467$ & $0.53592$ &
 $2.03195$ & $0.22240$ \\
$7$ &
 $~$ & $~$ &
 $5.72441$ & $1.15464$ &
 $3.07255$ & $0.39593$ \\
$8$ &
 $~$ & $~$ &
 $5.60990$ & $2.89864$ &
 $4.10682$ & $0.71922$ \\
$9$ &
 $~$ & $~$ &
 $~$ & $~$ &
 $4.83948$ & $1.37549$ \\
$10$ &
 $~$ & $~$ &
 $~$ & $~$ &
 $4.94943$ & $3.13117$ \\
\hline
$\sum_m a^{\luv}_m=$ &
 \multicolumn{2}{|c|}{$20.410$} &
 \multicolumn{2}{|c|}{$21.123$} &
 \multicolumn{2}{|c|}{$21.225$} \\
badness~$=$ &
 \multicolumn{2}{|c|}{$1.40\times 10^{-4}$} &
 \multicolumn{2}{|c|}{$8.42\times 10^{-6}$} &
 \multicolumn{2}{|c|}{$1.44\times 10^{-6}$} \\
\end{tabular}

\caption{Same as \tablename~\ref{tab:exp} but for the lux and luv
  profiles.\label{tab:lux}}
\end{table}

\clearpage

\clearpage
\begin{figure}
\includegraphics[width=\figurewidth]{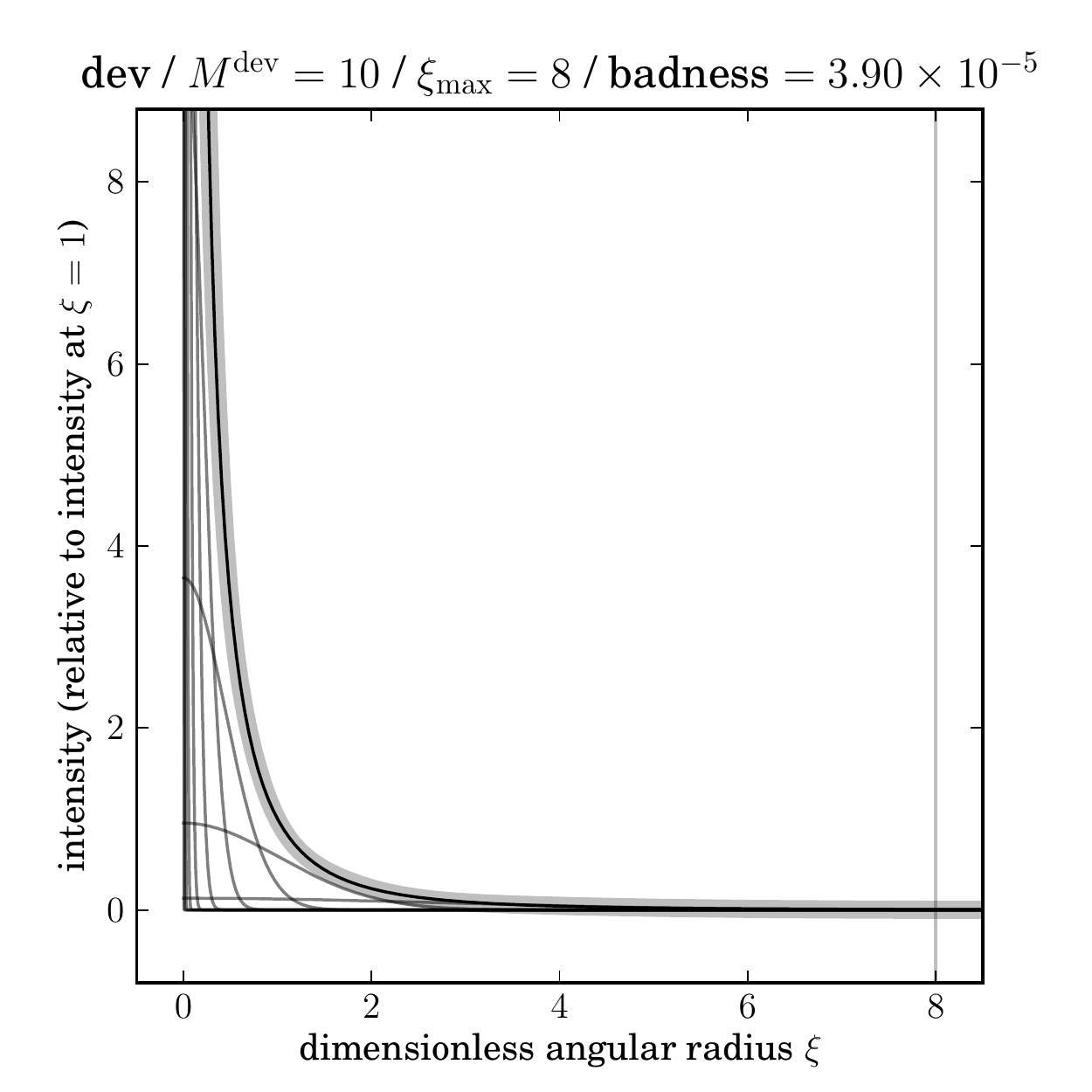}%
\includegraphics[width=\figurewidth]{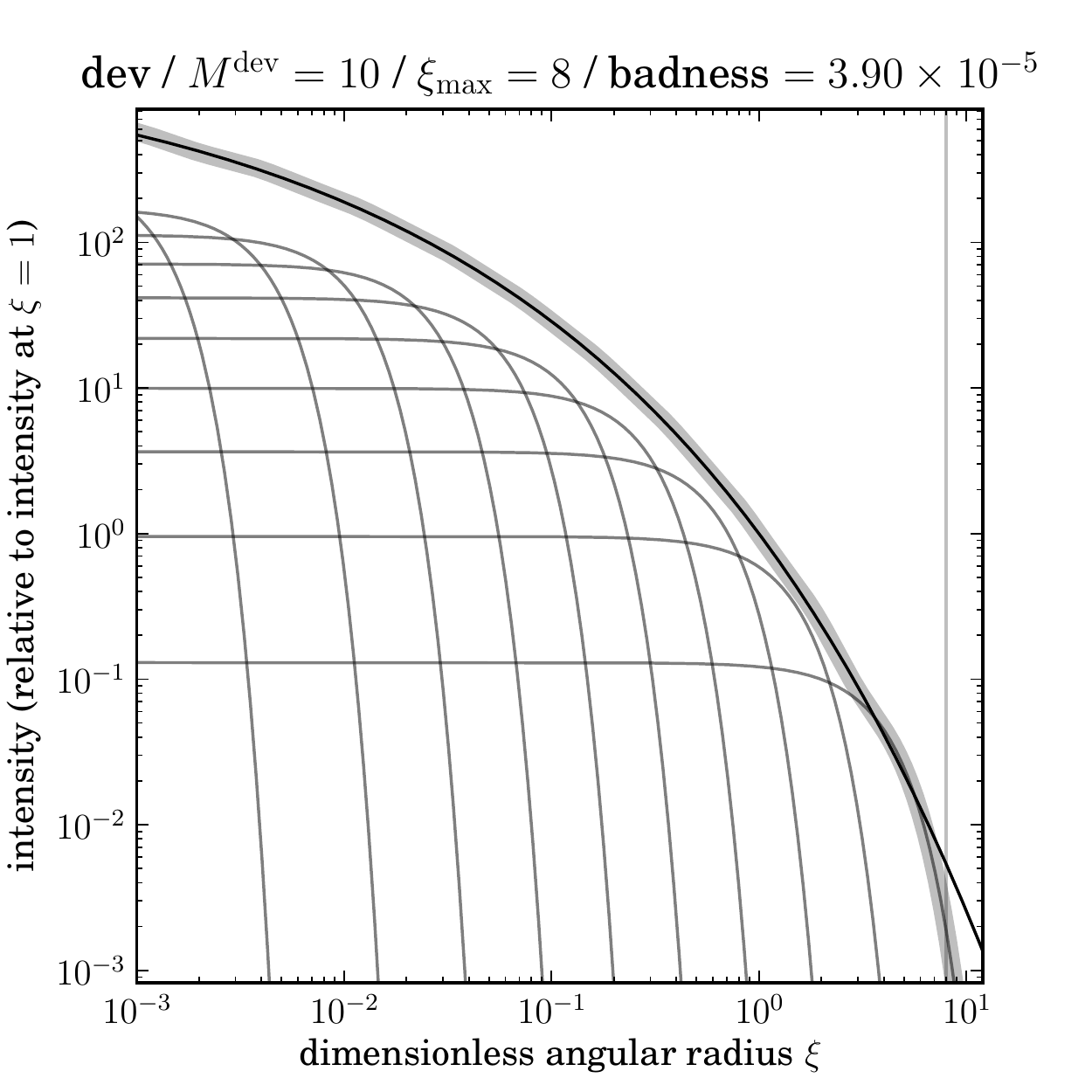}\\
\includegraphics[width=\figurewidth]{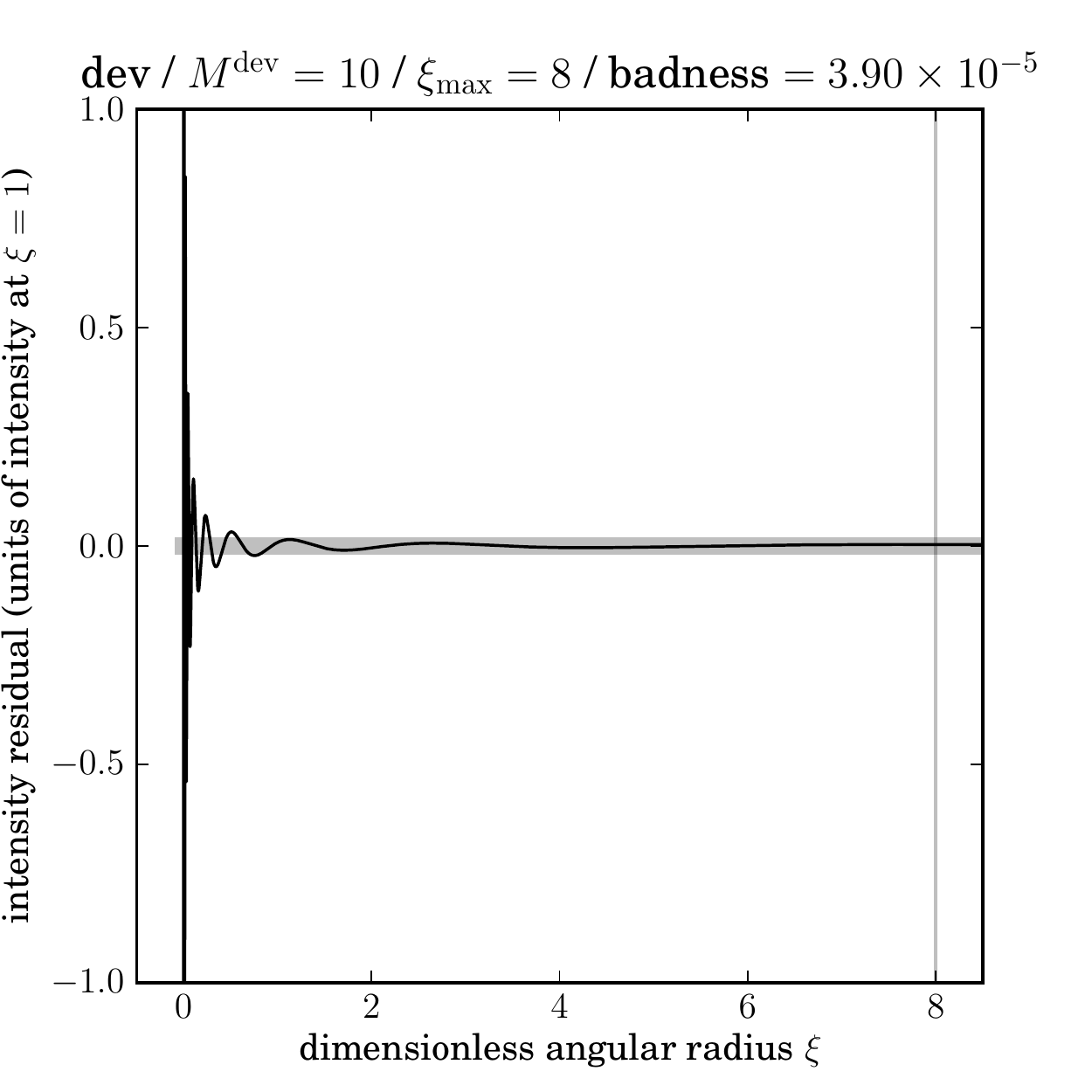}%
\includegraphics[width=\figurewidth]{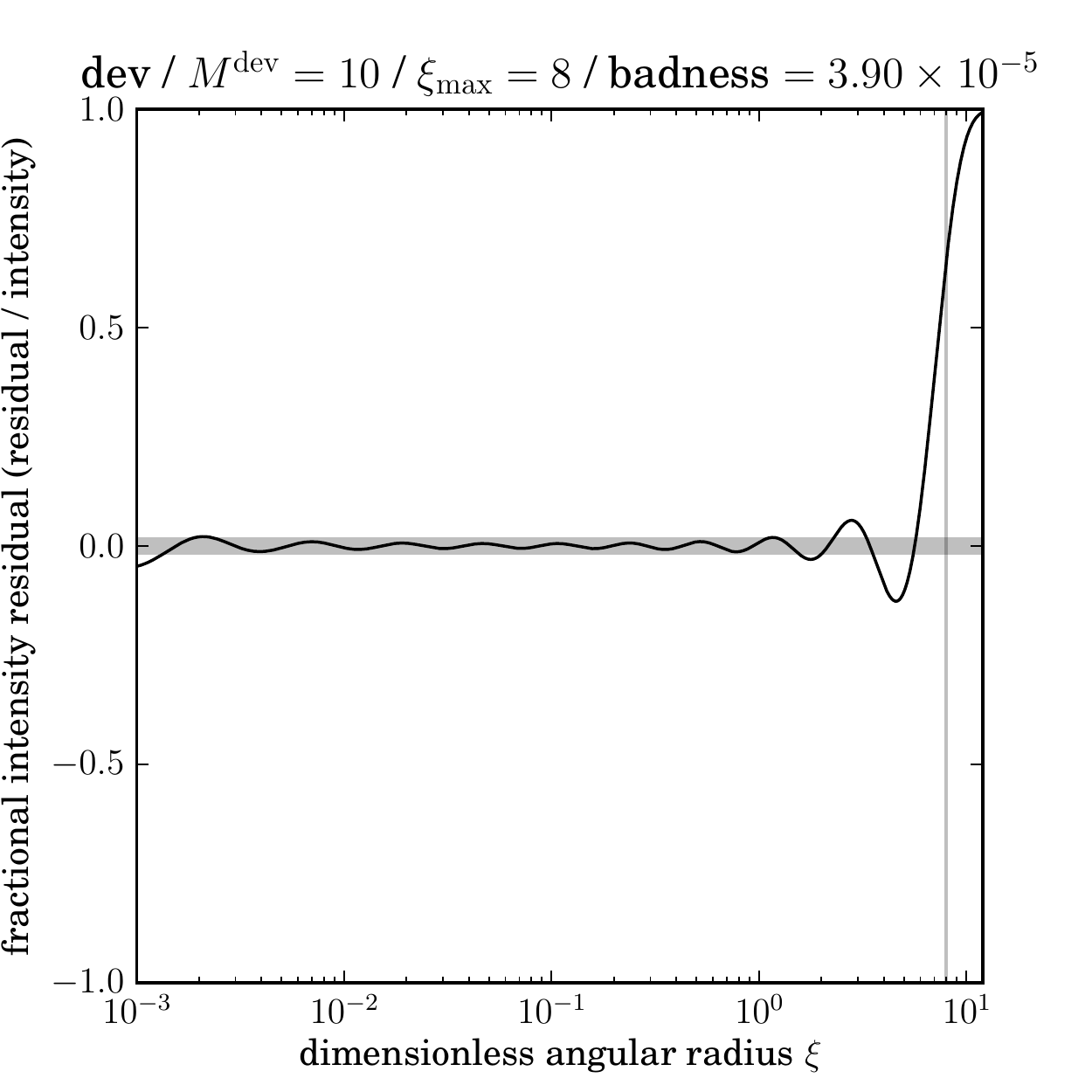}
\caption{The dev profile and the best $M^{\dev}=10$ approximation.
  The panels are equivalent to those in \figurename~\ref{fig:exp}.}
\end{figure}

\clearpage
\begin{figure}
\includegraphics[width=\figurewidth]{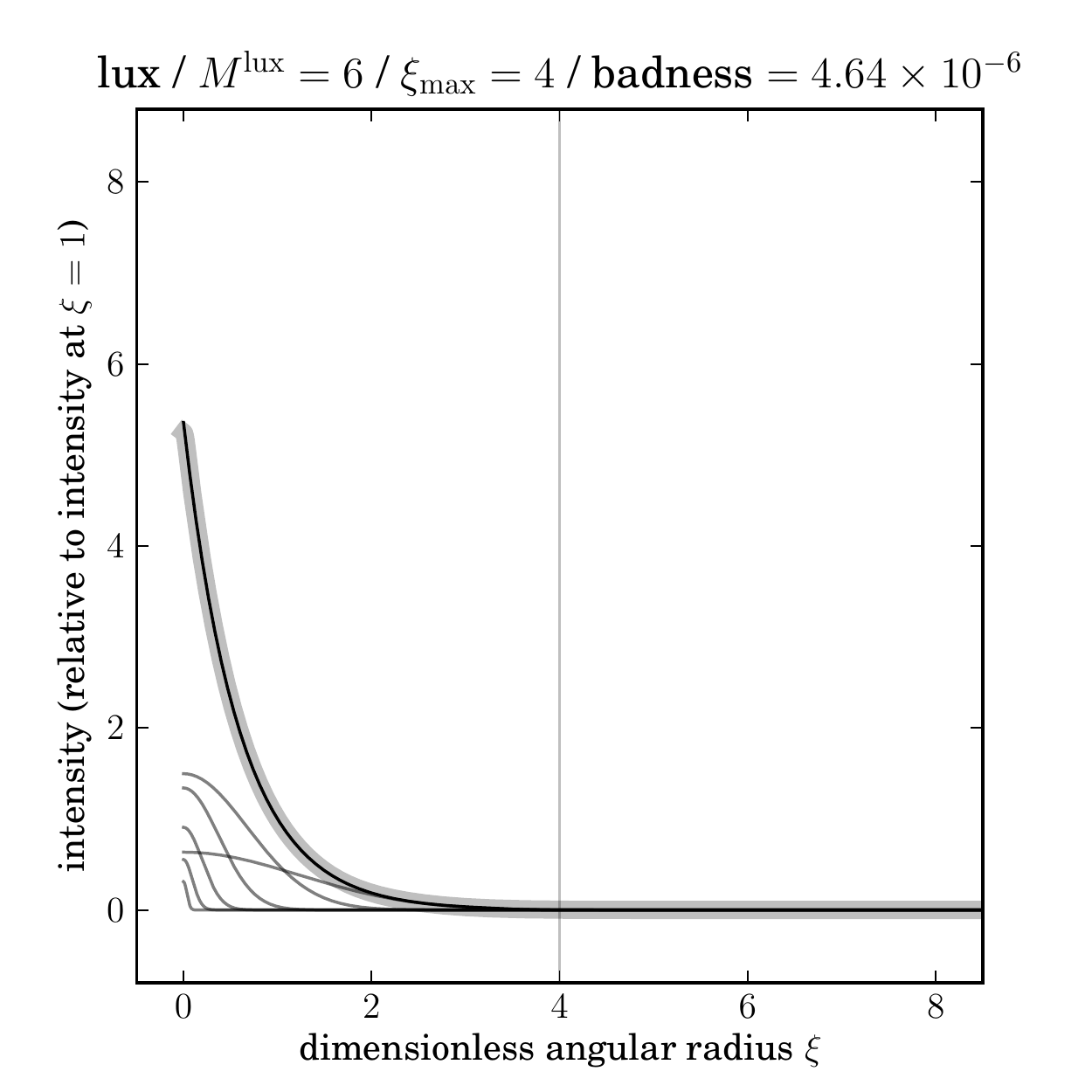}%
\includegraphics[width=\figurewidth]{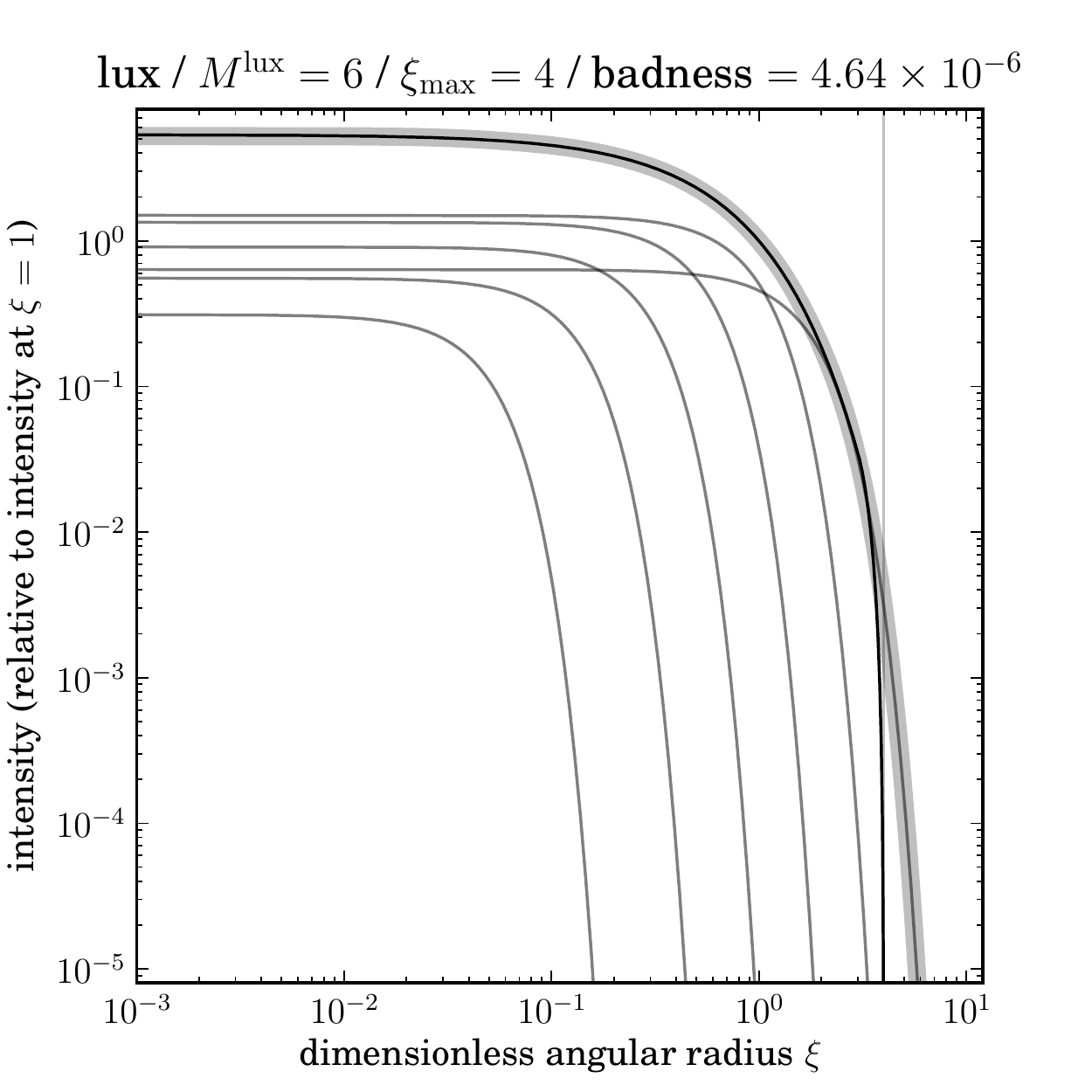}\\
\includegraphics[width=\figurewidth]{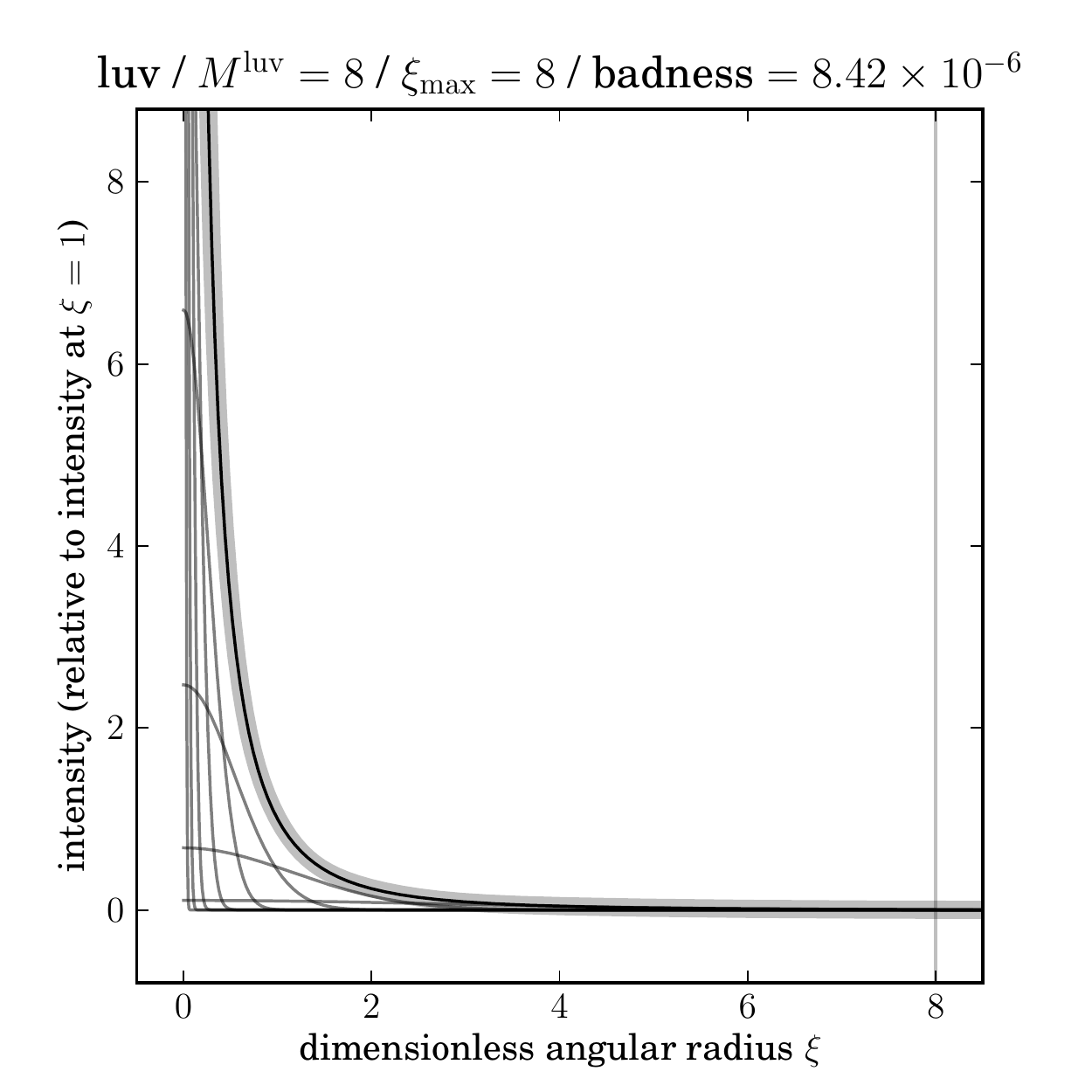}%
\includegraphics[width=\figurewidth]{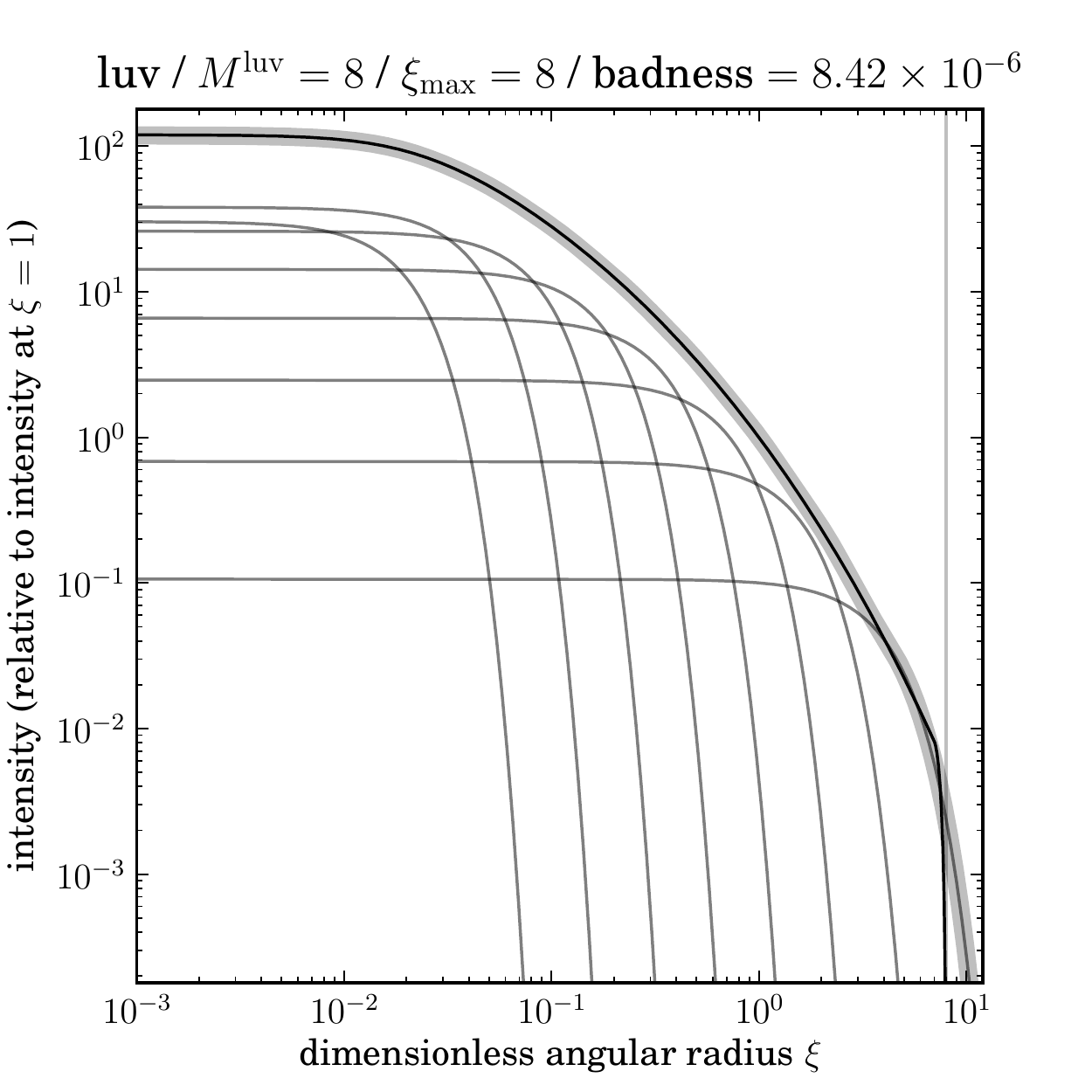}
\caption{The lux and luv profiles and approximations.  The panels are
  equivalent to those in the top-row of \figurename~\ref{fig:exp}.}
\end{figure}

\clearpage

\clearpage

\clearpage

\end{document}